\documentclass{article}

\usepackage[dvipsnames]{xcolor}
\definecolor{mygray}{gray}{0.6}

\usepackage[aboveskip=1pt,labelfont=bf,labelsep=period,justification=raggedright,singlelinecheck=off]{caption}

\usepackage{hyperref}
\hypersetup{colorlinks=true,citecolor=mygray,linkcolor=mygray,urlcolor=cyan}

\usepackage{arxiv}
\usepackage{amsmath,amssymb}
\usepackage[utf8]{inputenc}
\usepackage[T1]{fontenc}
\usepackage{url}
\usepackage{booktabs}
\usepackage{amsfonts}
\usepackage{nicefrac}
\usepackage{lipsum}
\usepackage{graphicx}
\usepackage[comma, sort&compress, numbers]{natbib}

\usepackage{siunitx}

\makeatletter
\newcommand{\thickhline}{%
  \noalign {\ifnum 0=`}\fi \hrule height 1.2pt
  \futurelet \reserved@a \@xhline
}
\makeatother

\usepackage{setspace}

\title{Modeling the influences of non-local connectomic projections on geometrically constrained cortical dynamics}

\author{ 
        {Rishikesan Maran\textsuperscript{1}, Eli J.~M\"uller\textsuperscript{1} and Ben D.~Fulcher\textsuperscript{1}} \\
        \\
        \textsuperscript{1}School of Physics, University of Sydney, Camperdown 2006, NSW, Australia.
}

\renewcommand{\shorttitle}{Modeling non-local influences on cortical dynamics}

\hypersetup{
pdftitle={Modeling the influences of non-local connectomic projections on geometrically constrained cortical dynamics},
pdfsubject={q-bio.NC, physics.bio-ph},
pdfauthor={Rishikesan Maran, Eli J.~M\"uller Ben D.~Fulcher},
pdfkeywords={neural field modeling, neural mass modeling, partial differential equations, systems biology},
}

\begin{document}
\maketitle

\begin{abstract}
The function and dynamics of the cortex are fundamentally shaped by the specific wiring configurations of its constituent axonal fibers, also known as the connectome.
However, many dynamical properties of macroscale cortical activity are well captured by instead describing the activity as propagating waves across the cortical surface, constrained only by the surface's two-dimensional geometry.
It thus remains an open question why the local geometry of the cortex can successfully capture macroscale cortical dynamics, despite neglecting the specificity of Fast-conducting, Non-local Projections (FNPs) which are known to mediate the rapid and non-local propagation of activity between remote neural populations.
Here we address this question by developing a novel mathematical model of macroscale cortical activity in which cortical populations interact both by a continuous sheet and by an additional set of FNPs wired independently of the sheet's geometry.
By simulating the model across a range of connectome topologies, external inputs, and timescales, we demonstrate that the addition of FNPs strongly shape the model dynamics of rapid, stimulus-evoked responses on fine millisecond timescales ($\lessapprox \SI{30}{\ms}$), but contribute relatively little to slower, spontaneous fluctuations over longer timescales ($> \SI{30}{ms}$), which increasingly resemble geometrically constrained dynamics without FNPs.
Our results suggest that the discrepant views regarding the relative contributions of local (geometric) and non-local (connectomic) cortico-cortical interactions are context-dependent: While FNPs specified by the connectome are needed to capture rapid communication between specific distant populations (as per the rapid processing of sensory inputs), they play a relatively minor role in shaping slower spontaneous fluctuations (as per resting-state functional magnetic resonance imaging).
\end{abstract}

\section{Introduction}
Cortical function emerges from complex interactions between spatially distributed neural populations, mediated by the structural axonal fibers that connect them \cite{bresslerLargescaleBrainNetworks2010}.
While the distribution of all fibers in the cortex is skewed towards shorter lengths \cite{horvatSpatialEmbeddingWiring2016}, some fibers are genetically encoded to connect specific remote populations \cite{thompsonGeneticInfluencesBrain2001, arnatkeviciuteGeneticInfluencesHub2021}, sometimes forming highly myelinated bundles \cite{schuzHumanCorticalWhite2002}.
Despite their substantial energy requirements for development and maintenance \cite{bullmoreEconomyBrainNetwork2012}, neurological experiments from the past century have demonstrated the critical role of these specifically positioned fibers---which we hereafter call fast-conducting non-local projections (FNPs)---in supporting cognition and behavior, by revealing the functional deficits that occur when they are damaged or removed \cite{sperryCerebralOrganizationBehavior1961, geschwindDISCONNEXIONSYNDROMESANIMALS1965, wernickeAphasischeSymptomenkomplex1974, goldsteinOrganismHolisticApproach1995, cataniRisesFallsDisconnection2005, oreillyCausalEffectDisconnection2013, filleyWhiteMatterCognition2016, graysonRhesusMonkeyConnectome2016}.
Stimulus--response experiments also demonstrate that FNPs are specialized to propagate macroscale cortical activity at speeds much faster than traveling waves on the continuous cortical surface, thus enabling rapid, non-local interactions between specific populations that bypass the constraints of the cortex's local geometry \cite{ferezouSpatiotemporalDynamicsCortical2007, mohajeraniSpontaneousCorticalActivity2013, limVivoLargeScaleCortical2012, liuCellClassspecificLongrange2024, hubatzSpatiotemporalPropertiesWhiskerevoked2020}.
Based on empirical observations including those above, FNPs are widely considered to be a core structural mechanism supporting the global integration of information in the brain \cite{parkStructuralFunctionalBrain2013, decoRareLongrangeCortical2021}.
Accordingly, several high-profile experimental efforts have aimed to develop comprehensive connectomes across species, which detail the specific populations that FNPs (as well as other fibers) connect \cite{ohMesoscaleConnectomeMouse2014, vanessenWUMinnHumanConnectome2013, kotterOnlineRetrievalProcessing2004}, and hence significantly accelerate our understanding of the structural underpinnings of cortical function and dysfunction \cite{bargmannConnectomeBrainFunction2013, fornitoConnectomicsNewParadigm2015}.

Despite extensive evidence demonstrating the key role of FNPs in supporting cortical function and shaping cortical dynamics, recent analyses have indicated that many dynamical properties of macroscale activity, on millimeter lengthscales and above, can be well captured by geometrically constrained models of neural activity that neglect the specificity of connectome-informed FNP connectivity \cite{robinsonDeterminationDynamicBrain2021, hendersonEmpiricalEstimationEigenmodes2022, pangGeometricConstraintsHuman2023}.
This body of work suggests that macroscale cortical activity is well approximated as the superposition of traveling waves on a physically continuous cortical sheet, such that the waves are constrained only by the local two-dimensional geometry of the cortical surface rather than the specific wiring positions of the cortex's constituent axonal fibers.
This geometric description of cortical activity is derived by a broad class of generative models called neural field models \cite{nunezNeocorticalDynamicsHuman1995, jirsaFieldTheoryElectromagnetic1996, jirsaDerivationMacroscopicField1997, robinsonPropagationStabilityWaves1997} which assume a mean approximation rule of cortical structural connectivity that is spatially homogeneous (exhibiting the same properties at all points) and spatially isotropic (exhibiting the same properties in all directions).
It remains an unresolved and fundamental problem in macroscale neuroscience how neural field models, which commonly assume homogeneous and isotropic FNP connectivity, in contrast to the reliable heterogeneity observed in the connectome \cite{markovRoleLongrangeConnections2013, betzelSpecificityRobustnessLongdistance2018, vandenheuvelRichClubOrganizationHuman2011, hendersonUsingGeometryUncover2013}, can nevertheless generate predictions of cortical dynamics that closely match many experimental observations.

Here we seek to mechanistically reconcile two seemingly conflicting empirical findings: (i) direct evidence of the role of FNPs in facilitating the non-local dynamics of stimulus-evoked whole-brain responses measured with voltage-sensitive dye (VSD) imaging \cite{ferezouSpatiotemporalDynamicsCortical2007, mohajeraniSpontaneousCorticalActivity2013, limVivoLargeScaleCortical2012}, electroencephalography (EEG) \cite{massiminiCorticalMechanismsLoss2012, seguinCommunicationDynamicsHuman2023}, and wide-field calcium imaging \cite{liuCellClassspecificLongrange2024}; and (ii) the predictive successes of geometrically constrained neural field models, which ignore the specificity of FNP connectivity informed by connectomic data, in capturing the spatiotemporal dynamics of resting-state functional magnetic resonance imaging (fMRI) \cite{robinsonDeterminationDynamicBrain2021, hendersonEmpiricalEstimationEigenmodes2022, pangGeometricConstraintsHuman2023}, the spatial patterns of fMRI task activation maps \cite{hendersonEmpiricalEstimationEigenmodes2022, pangGeometricConstraintsHuman2023}, and resting-state magnetoencephalography (MEG) \cite{hendersonEmpiricalEstimationEigenmodes2022}.
To this end, we develop a new mathematical model of macroscale cortical activity in which interactions between cortical populations are mediated both by: (i) a continuous sheet of fibers that are homogeneously and isotropically distributed in space; and also by (ii) a set of specific FNPs whose wiring positions are independent of the geometry of the sheet.
Our phenomenological model is a simple extension of a neural field model, in which the additionally wired FNPs perturb the dynamics of the traveling waves along the cortical surface by behaving as `topological shortcuts' that facilitate rapid propagation between specific distant populations (at speeds faster than the traveling waves).
By simulating the model over a variety of FNP connectome topologies, we find that the influence of FNPs on the model dynamics is strongly dependent on two key factors: (i) the spatial precision of the external driving input (e.g., stimulus-evoked responses versus spontaneous activity with no explicit external stimulus); and (ii) the timescales over which spatiotemporal activity is resolved (e.g., the subsecond timescales of VSD versus the second timescales of fMRI).
Our results thus provide a plausible explanation for the empirical findings above, demonstrating that the connectome (which informs the specificity of FNP connectivity) and cortical geometry (which spatially aggregates this specificity) are best suited to describe and capture macroscale cortical dynamics from contrasting experimental settings and modalities.

\section{Materials and Methods}

In order to investigate the additional contribution of fast non-local interactions to geometrically constrained cortical dynamics, we developed a mathematical model of macroscale cortical population-level activity (on millimeter lengthscales and above).
Our model isolates the role of specific arrangements of FNPs in perturbing cortical dynamics, separately from the influence of the sheet geometry.
The model's structure is a two-dimensional continuous sheet representing the cortical surface, augmented with a set of FNPs that connect pairs of points on the sheet.
While there are existing models that combine geometric and connectomic mechanisms of interaction \cite{spieglerSilicoExplorationMouse2020, spieglerSelectiveActivationRestingState2016, qubbajNeuralFieldDynamics2007, jirsaNeuralFieldDynamics2009, jirsaSpatiotemporalPatternFormation2000}, the model introduced here supports finite-speed traveling waves on a two-dimensional cortical sheet with multiple FNPs, thereby overcoming key limitations that render prior models unsuited to our investigations (discussed further in Discussion).

\subsection*{Model Description}

\begin{figure}[!ht]
    \centering
    \includegraphics[width = \textwidth]{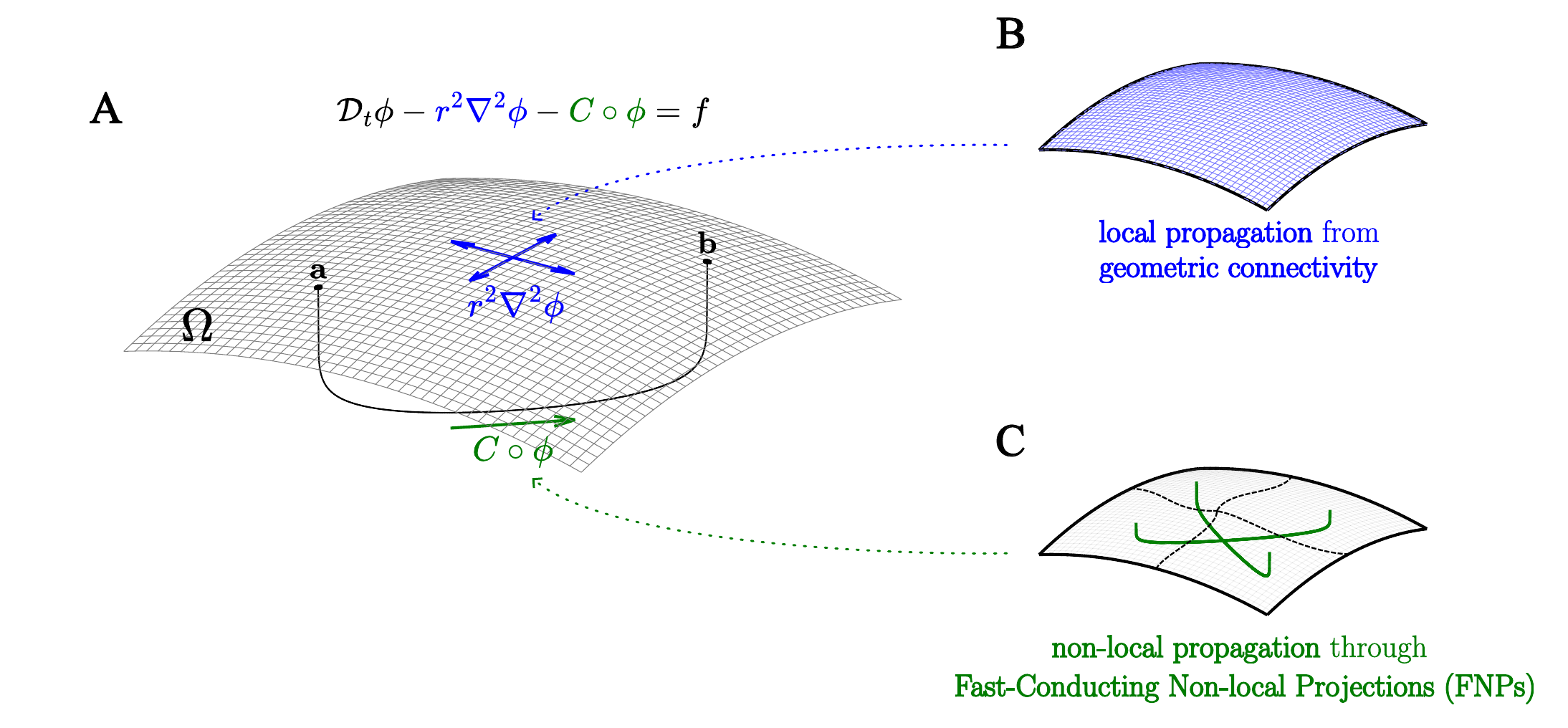}
    \captionsetup{font=small}
    \caption{
    \textbf{Schematic of our mathematical model of macroscale cortical activity with local geometric (blue) and non-local connectomic (green) mechanisms of cortico-cortical interaction.}
    The model equations are defined in Eq.~\eqref{eq:model}.
    \textbf{A.} 
    The two-dimensional continuous surface $\Omega$ represents the cortical sheet of a cerebral hemisphere, on which neural populations are represented as points.
    The mean population-level activity at each point, on millimeter lengthscales and above, is described by variable $\phi$. 
    This model includes both local non-local mechanisms of interaction between points on $\Omega$.
    Local geometric propagation of activity takes the form of traveling waves along $\Omega$, while non-local connectomic propagation takes the form of rapid, non-local interactions between spatially distant pairs of points on $\Omega$ (an example of which is illustrated as $\mathbf{a}$ and $\mathbf{b}$).
    \textbf{B.} Geometric propagation (traveling waves) is derived from a spatially homogeneous and isotropic connectivity rule \cite{jirsaFieldTheoryElectromagnetic1996, jirsaDerivationMacroscopicField1997, robinsonPropagationStabilityWaves1997}.
    \textbf{C.} Connectomic propagation between specific pairs of points is facilitated by fast-conducting, non-local projections (FNPs), which are axonal fibers that propagate activity faster than traveling waves along $\Omega$, and is considered to underpin macroscale cortical function by facilitating rapid inter-communication between spatially distributed neural systems.
    FNPs introduce spatial inhomogeneities and anisotropies to the otherwise geometric connectivity.
    }
    \label{fig:schematic}
\end{figure}

Figure~\ref{fig:schematic}A illustrates the architecture of our model---containing both geometrically constrained and fast non-local connectomic mechanisms of cortico-cortical interaction---which we describe below.
The cortical sheet of a cerebral hemisphere is represented as a continuous two-dimensional surface, denoted $\Omega$.
Cortical populations are identified by their spatial location $\mathbf{r}$ on $\Omega$, and their neural activity over time is represented by the field variable $\phi(\mathbf{r}, t)$. 
The spatiotemporal evolution of $\phi(\mathbf{r}, t)$ over time $t \in [0, \infty)$ is governed by the non-local partial differential equation:
\begin{align} 
    &\mathcal{D}_t \phi - r^2 \nabla^2 \phi - C \circ \phi = f\,,  \quad \text{ for all } (\mathbf{r}, t) \in \Omega \times [0, \infty)\,; \label{eq:model} \\
    & \phi(\mathbf{r}, 0) = \dot \phi(\mathbf{r}, 0) = 0 \,. \label{eq:initconditions}
\end{align}
In Eq.~\eqref{eq:model}, $\mathcal{D}_t$ is a second-order linear differential operator in time, and the term $r^2 \nabla^2 \phi(\mathbf{r}, t)$ captures local spatial relations on $\Omega$ for different points $\mathbf{r}$.
Both terms together capture local propagation as traveling waves along $\Omega$, which arises from a spatially homogeneous and isotropic connectivity rule (illustrated in Fig.~\ref{fig:schematic}B) \cite{jirsaFieldTheoryElectromagnetic1996, jirsaDerivationMacroscopicField1997, robinsonPropagationStabilityWaves1997}.
The term $(C \circ \phi)(\mathbf{r}, t)$ captures non-local propagation between specific points on $\Omega$ facilitated by FNPs (forming `shortcuts' through $\Omega$, as illustrated in Fig.~\ref{fig:schematic}C).
Each FNP is a unidirectional projection from an assigned source point (denoted `$\mathbf{a}$' in Fig.~\ref{fig:schematic}A) to a target point (denoted `$\mathbf{b}$' in Fig.~\ref{fig:schematic}A), through which activity propagates more rapidly than by a traveling wave along $\Omega$.
The term $f(\mathbf{r}, t)$ is the external input that drives the model's activity.
In the absence of input, $f \equiv 0$, the initial conditions in Eq.~\eqref{eq:initconditions} define zero activity at all points $\mathbf{r}$ and for all time $t$.
Further details on the above terms, and a derivation of Eq.~\eqref{eq:model} from physiological principles, are in Sec.~\nameref{SI:modelderivation}.

A notable property of the non-local term $C\circ\phi$ is that, for any given input $f$, the total activity in the system integrated over space and time is preserved after the addition of any FNP to the structural connectivity.
This property prevents non-oscillatory instabilities in the dynamics, which can occur when activity is regenerated from FNPs at a faster rate than activity globally dissipates \cite{qubbajNeuralFieldDynamics2007, jirsaNeuralFieldDynamics2009}, and therefore allows multiple FNPs to be added with arbitrary connectivity strengths.
In alternative models, total activity is not preserved (e.g., the neural field model by Jirsa and colleagues \cite{qubbajNeuralFieldDynamics2007, jirsaNeuralFieldDynamics2009, jirsaSpatiotemporalPatternFormation2000}), such that the connectivity strengths of each FNP must be progressively reduced to maintain linear stability.

The model defined by Eq.~\eqref{eq:model} will be used to simulate cortical dynamics for different external inputs $f(\mathbf{r}, t)$, so that the perturbatory influence of FNPs on the dynamics can be evaluated separately for each input setting.
For example, to simulate stimulus-evoked dynamics in response to an impulse stimulus, $f$ is set as an impulse-like input that is Gaussian in both space and time:\cite{kerrPhysiologybasedModelingCortical2008, muktaNeuralFieldTheory2019}:
\begin{align}
    &f^\text{stim}(\mathbf{r}, t) \propto \exp\left(\frac{-\lVert\mathbf{r} -\mathbf{r}_{os}\rVert^2}{2 \sigma_x^2} \right) \exp\left(\frac{-(t -t_{os})^2}{2 \sigma_t^2} \right); \nonumber \\ &\int_0^\infty dt \int_\Omega d^2\mathbf{r} \ f^\text{stim} (\mathbf{r}, t) = 1\,; \label{eq:impulsestimulus}
\end{align}
where $\mathbf{r}_{os} \in \Omega, t_{os} > 0$ are the position and onset time of the stimulus, respectively, and $\sigma_x, \sigma_t$ are the spatial and temporal width of the stimulus, respectively.
In contrast, to simulate noise-driven dynamics---that is, temporally unstructured inputs---we set $f$ as a continuous space--time stochastic process:
\begin{align}
    &f^\text{noise}(\mathbf{r}, t) = g(\mathbf{r}) \dot W(\mathbf{r}, t)\,; \nonumber \\
    &\mathbb{E}(\dot W(\mathbf{r}, t) \dot W(\mathbf{r}', t')) = \delta(\mathbf{r} - \mathbf{r}') \delta(t - t')\,; \label{eq:whitenoise}
\end{align}
where $\dot W$ on $\Omega \times [0, \infty)$ is a space-time white noise process \cite{sanz-leonNFTsimTheorySimulation2018, robinsonDeterminationDynamicBrain2021}, and $g(\mathbf{r})$ on $\Omega$ is a smooth deterministic function which controls the variance of the process at different points on $\Omega$.

\subsection*{Model Implementation}

To numerically solve the model defined by Eq.~\eqref{eq:model}, we developed a finite-difference scheme used for non-local partial differential equations, in which derivatives in time and space are approximated by centered finite differences and definite integrals are approximated using a quadrature rule \cite{kavallarisNonLocalPartialDifferential2018}.
For our investigations, $\Omega$ is taken to be a square with periodic boundary conditions, making its geometry equivalent to a flat torus.
A torus has a closed surface, like a more physically realistic sphere \cite{gabayCorticalGeometryDeterminant2017}, but permits more efficient numerical simulations on a square grid.
The side length of the square was set to $L = \SI{0.4}{m}$, so that the area of $\Omega$ is approximately that of a human cortical hemisphere \cite{robinsonPropagationStabilityWaves1997}.
We then uniformly partitioned a grid of points from $\Omega$ with grid spacing $\Delta x$, and time points from $[0, \infty)$ with timestep $\Delta t$.
With this time and space partitioning, we used a finite-difference scheme to numerically solve Eq.~\eqref{eq:model}.
Further details and the derivation of this scheme, including the chosen grid spacing and timestep for our investigations, can be found in Sec.~\nameref{SI:NumericalTreatment}.

The parameters of Eq.~\eqref{eq:model} (detailed in Eqs~\eqref{SI:eq:modeldetailed},~\eqref{SI:eq:differentialoperator},~\eqref{SI:eq:nonlocalterm} in Sec.~\nameref{SI:modelderivation}) and their values used in our investigations are given in Table~\ref{tab:implementation}.
For clarity, we divide the parameters into three groups:
(i) parameters that govern geometric propagation through traveling waves: $r$, $\gamma$, $\nu_0$;
(ii) parameters that govern interaction via FNPs: $\{c_m\}_m$, $\{\tau_m\}_m$; and
(iii) parameters that govern the spatiotemporal properties of the external input $f$: $\sigma_x, \sigma_t$.
The value of parameter $r$, the characteristic lengthscale of the mean connectivity rule, and the value of $\gamma$, the traveling wave speed divided by $r$, were both taken from the two-dimensional neural field model developed by Robinson et al. for the human cortex \cite{robinsonPropagationStabilityWaves1997, robinsonEstimationMultiscaleNeurophysiologic2004}.
The value of $\nu_0$, which is the effective rate at which activity at a given position is regenerated, was taken from the excitatory-only neural field model developed by Jirsa and Haken \cite{jirsaDerivationMacroscopicField1997}.

\begin{table}[!ht]
\centering
\caption{
{\bf Nominal model parameter values for numerical experiments presented here.}
}
\begin{tabular}{|l|l|l|}
\hline
\bf{Parameter} & \bf{Description} & \bf{Value} \\ \thickhline
\multicolumn{3}{|l|}{Topology} \\ \hline
$L$ & Length of $\Omega$ (as a square) & $\SI{0.4}{m}$ \\ \hline
\multicolumn{3}{|l|}{Local geometric Propagation} \\ \hline
$r$ & lengthscale of mean connectivity & $\SI{0.086}{m}$ \\ \hline
$\gamma$ & Wave speed divided by $r$ & $\SI{116}{s^{-1}}$\\ \hline
$\nu_0$ & Cortical Gain & $\SI{0.756}{}$ \\ \hline
\multicolumn{3}{|l|}{Non-local Connectomic Propagation} \\ \hline
$c_m$ & Connectivity strength of $m^\text{th}$ FNP & $\SI{7.396e-3}{m^2}$ for all $m$ \\ \hline
$\tau_m$ & Conduction time of $m^\text{th}$ FNP & $\SI{0}{s}$ for all $m$ \\ \hline
\multicolumn{3}{|l|}{Input} \\ \hline
$\sigma_x$ & Spatial width of stimulus input & $\SI{4e-3}{m}$ \\ \hline
$\sigma_t$ & Temporal width of stimulus input & $\SI{6e-4}{s}$ \\ \hline
\end{tabular}
\label{tab:implementation}
\end{table}

We set $c_m$, which is the connectivity strength of the $m^\text{th}$ FNP, to equate the strength of local and non-local propagation from the source of the FNP.
Accordingly, we set $c_m = r^2$ for each $m$, so that when the model is numerically treated, the connectivity strength of the FNP ($c_m/(\Delta x)^2$ in Eq.~\eqref{SI:eq:laplaciannumericaltreatment} in Sec.~\nameref{SI:NumericalTreatment}) is equal to the strength of coupling between geometrically nearest neighbors on the grid from the local geometry ($r^2/(\Delta x)^2$ in Eq.~\eqref{SI:eq:nonlocalnumericaltreatment} in Sec.~\nameref{SI:NumericalTreatment}).
We set the value of $\tau_m$, which represents the conduction time of the $m$th FNP, so that it is always smaller than the conduction time of a traveling wave between the FNP source and target, but is also small enough to avoid oscillatory instabilities independently of the chosen connectome configuration \cite{qubbajNeuralFieldDynamics2007, jirsaNeuralFieldDynamics2009}.
However, the conduction speeds of FNPs vary widely from factors other than their length \cite{swadlowAxonalConductionDelays2012}, and the critical values of the $\tau_m$ values separating stability from instability depends on the spatial configurations of the FNPs \cite{qubbajNeuralFieldDynamics2007}.
For this reason, we fixed $\tau_m = 0$ for each $m$.
The values of $\sigma_x, \sigma_t$ (in Eq.~\eqref{eq:impulsestimulus}), the spatial and temporal widths of the impulse stimulus, respectively, were set to sufficiently small values such that the input mimics a delta impulse function without causing spurious oscillations in the numerical scheme.

\section{Results}

Using the phenomenological model defined in Eq.~\eqref{eq:model}, we aim to shed light on the variable role of FNPs across different experimental settings, with specific individual FNPs playing a clear role in facilitating corticocortical interactions that underpin distributed information processing in some settings \cite{cataniRisesFallsDisconnection2005, filleyWhiteMatterCognition2016, ferezouSpatiotemporalDynamicsCortical2007, mohajeraniSpontaneousCorticalActivity2013}, but being apparently negligible relative to idealized geometric connectivity rules in others \cite{robinsonDeterminationDynamicBrain2021, hendersonEmpiricalEstimationEigenmodes2022, pangGeometricConstraintsHuman2023}.

\subsection*{Model Demonstration}

\begin{figure}[!ht]
    \centering
    \includegraphics[width = \textwidth]{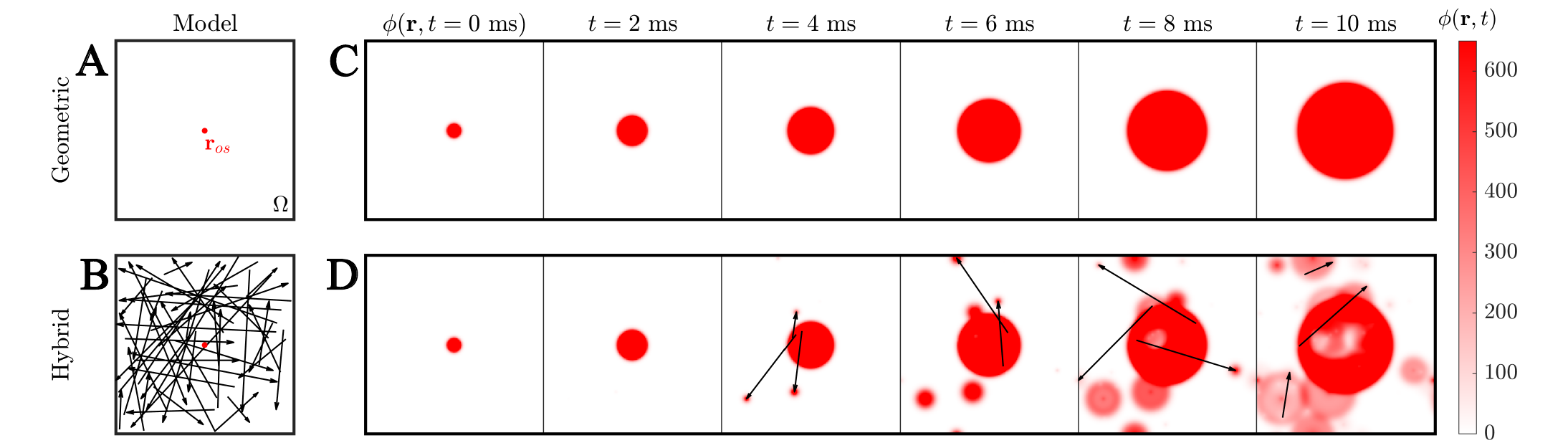}
    \captionsetup{font=small}
    \caption{\textbf{FNPs shape stimulus-evoked dynamics by enabling rapid, non-local propagation of activity between neural populations.}
    We used the model defined in Eq.~\eqref{eq:model} to examine the influence of FNPs on stimulus-evoked dynamics, by comparing the model's stimulus-evoked dynamics under two structural connectivity configurations, in response to the same impulse stimulus.
    \textbf{A.} The geometric model is governed by a spatially homogeneous and isotropic connectivity rule, and is purely constrained by the geometry of $\Omega$ (shown as the square).
    \textbf{B.} The hybrid model's structural connectivity additionally contains 50 FNPs, depicted as arrows from their source to target.
    The impulse stimulus applied to both models was positioned at the point $\mathbf{r}_{os}$, annotated with a red dot.
    \textbf{C.} The geometric model's stimulus-evoked response is illustrated as heatmaps of activity profile $\phi(\mathbf{r}, t)$ at time points, $t = 0,2,4,6,8,10$\,ms, measured relative to stimulus onset.
    \textbf{D.} The hybrid model's stimulus-evoked response is illustrated in a similar manner as in \textbf{C}.
    Specific FNPs of the hybrid connectivity (depicted in \textbf{B}) facilitate rapid activity propagation from their source to target at different times and are annotated manually in each heatmap.
    Note that we have imposed an upper limit of $\phi = 650$ on the color scale for all heatmaps in \textbf{C} and \textbf{D}, to facilitate clear visibility across all time points.
    }
    \label{fig:proofofprinciple}
\end{figure}

We first introduce the behavior of the model, through a simple demonstration of how the incorporation of FNPs to the model's structural connectivity perturbs its stimulus-evoked dynamics (i.e., evoked response to an impulse stimulus, cf. Eq.~\eqref{eq:impulsestimulus}).
Two configurations of structural connectivity were considered.
The geometric model, depicted in Fig.~\ref{fig:proofofprinciple}A, is governed by the spatially homogeneous and isotropic structural connectivity rule of neural field models, thus completely constrained by the geometry of $\Omega$ and absent of any FNPs \cite{robinsonPropagationStabilityWaves1997}.
The hybrid model, depicted in Fig.~\ref{fig:proofofprinciple}B, additionally incorporates 50 FNPs whose source and target positions were each sampled uniformly at random from $\Omega$.
We then computed the response of both models to an impulse stimulus at position $\mathbf{r}_{os} = (L/2,L/2)$, shown with a red dot in Figs~\ref{fig:proofofprinciple}A and B.
The stimulus-evoked responses of the geometric and hybrid models are plotted in Figs~\ref{fig:proofofprinciple}C and D, respectively, as heatmaps of their spatial activity profiles $\phi(\mathbf{r}, t)$, at each of six time points following stimulus onset.
Consistent with the connectivity rule of neural field models, the stimulus-evoked response of the geometric model is a traveling wave emanating from $\mathbf{r}_{os}$.
By contrast, the response of the hybrid model (containing FNPs) also comprises the activation of specific non-local points due to non-local activity propagation via FNPs.
These non-local activations lead to additional traveling wave fronts under the geometric connectivity, yielding dynamics comprising a superposition of multiple traveling waves.
In summary, FNPs yield more complex stimulus-evoked dynamics by non-locally propagating activity between spatially distant populations at speeds faster than the traveling waves constrained by cortical geometry.

\subsection*{Influence of a single FNP on cortical dynamics resolved across varying timescales}

We aim to understand how a single FNP's perturbation on the model's stimulus-evoked dynamics evolves over different timescales.
We then use this distinction of perturbation by timescale to shed light on how the influence of FNPs on measured macroscale cortical dynamics could differ between activity resolved over fast sub-second timescales by modalities such as voltage-sensitive dye (VSD) imaging, EEG or calcium imaging \cite{ferezouSpatiotemporalDynamicsCortical2007, mohajeraniSpontaneousCorticalActivity2013, limVivoLargeScaleCortical2012, massiminiCorticalMechanismsLoss2012, seguinCommunicationDynamicsHuman2023, liuCellClassspecificLongrange2024}; and activity resolved over slower, second timescales by modalities such as fMRI \cite{robinsonDeterminationDynamicBrain2021, hendersonEmpiricalEstimationEigenmodes2022, pangGeometricConstraintsHuman2023}.

\begin{figure}[!ht]
    \centering
    {\includegraphics[width = \textwidth]{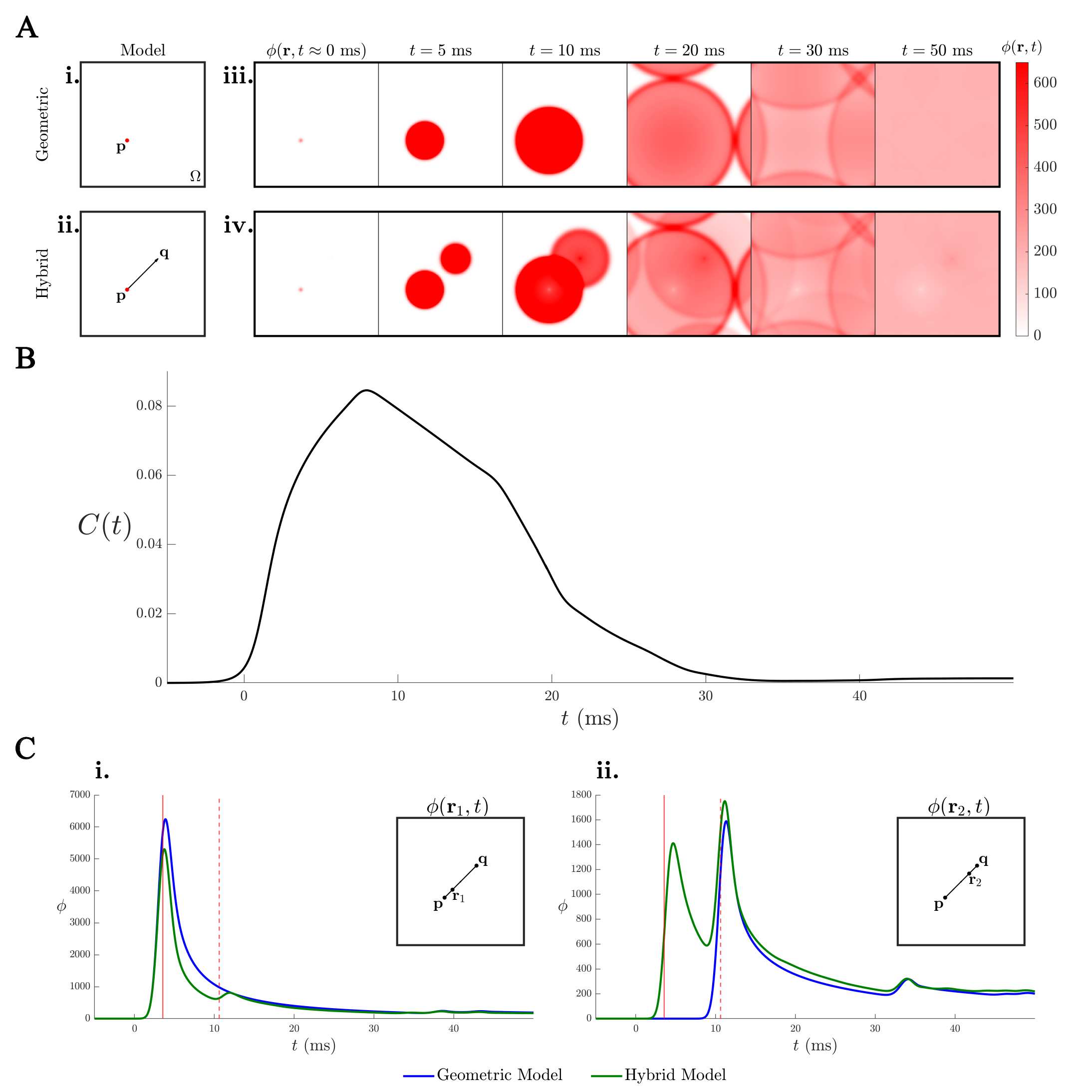}}\qquad
    \captionsetup{font=small}
    \caption{
    \textbf{The addition of a single fast-conducting non-local projection (FNP) predominantly shapes stimulus-evoked dynamics on short timescales $\lessapprox \SI{30}{\ms}$.}
    \textbf{A.} We compared the stimulus-evoked dynamics of: \textbf{i.} the geometric model, and \textbf{ii.} a hybrid model whose structural connectivity additionally contains a single FNP from $\mathbf{p} = (3L/8, 3L/8)$ to $\mathbf{q} = (5L/8,5L/8)$.
    An impulse stimulus was applied to both models at position $\mathbf{p}$, shown by the red dot. 
    \textbf{iii.} and \textbf{iv.} show heatmaps of the stimulus-response dynamics $\phi(\mathbf{r},t)$ of the geometric and hybrid models for $t = 0, 5, 10, 20, 30, 50$\,ms.
    Note that the first snapshot, labeled $t \approx 0$, was taken at a time slightly earlier than stimulus onset ($-\SI{1.33}{\ms}$), to indicate that $\mathbf{q}$ is activated after $\mathbf{p}$ in the hybrid dynamics.
    The color scale for all heatmaps is shown on the right of the figure.
    \textbf{B.} The cosine dissimilarity (or cosine distance) between the spatial activity patterns of the geometric and hybrid model's stimulus-evoked dynamics over time, denoted $C(t)$ (Eq.~\eqref{SI:eq:cosinedissimilarity} in Sec.~\nameref{SI:ResultsDetails}), quantifies the FNP's perturbation on the model's stimulus-evoked dynamics over time.
    \textbf{C.} We examined the difference between the geometric and hybrid model's stimulus-evoked response at two locations: \textbf{i.} $\mathbf{r}_1 = (7L/16,7L/16)$, and \textbf{ii.} $\mathbf{r}_2 = (9L/16,9L/16)$.
    Insets in the top right corner of each plot show the positions of $\mathbf{r}_1$ and $\mathbf{r}_2$.
    The solid and dashed red lines in both plots represent the time at which the perturbation of the FNP on the dynamics commences $(t \approx \SI{4}{\ms})$ and ends $(t \approx \SI{11}{\ms})$, respectively.
    }
    \label{fig:timescales}
\end{figure}

We compared the stimulus-evoked response of the geometric model depicted in Fig.~\ref{fig:timescales}A(i), and a hybrid model containing a single additional FNP from $\mathbf{p} \rightarrow \mathbf{q}$, depicted in Fig.~\ref{fig:timescales}A(ii).
The length of this FNP ($L\sqrt{2}/4 \approx \SI{14}{\cm}$) is comparable to the longest intra-hemispheric FNPs measured in human cortex \cite{robertsContributionGeometryHuman2016}.
The impulse stimulus applied to both model was set to position $\mathbf{p}$, shown as a red dot in Figs~\ref{fig:timescales}A(i) and (ii), which was selected to ensure that the FNP $\mathbf{p}\to\mathbf{q}$ would propagate activity from $\mathbf{p}$ to $\mathbf{q}$ immediately at stimulus onset.
The stimulus-evoked responses of both the geometric and hybrid model to this impulse stimulus are shown in Figs~\ref{fig:timescales}A(iii) and (iv).
As expected, the FNP perturbs the stimulus-evoked dynamics by non-locally activating $\mathbf{q}$, which results in an additional traveling wave emanating from $\mathbf{q}$.
However, from the point that the two traveling waves begin to superimpose, from $t \approx \SI{10}{\ms}$ onwards, the geometric and hybrid model's response becomes progressively similar.
At the later time points shown in Fig.~\ref{fig:timescales}A, $t = 30, \SI{50}{\ms}$, we see that the two models exhibit very similar spatial patterns in their responses, with the hybrid model's spatial activity pattern displaying a very small perturbation, near $\mathbf{p}$ and $\mathbf{q}$, due to the perpetual propagation of activity between these two points via the FNP.
To quantify the timescale dependency of the FNP's perturbation on the dynamics, we measured the cosine dissimilarity metric (or cosine distance) between the spatial activity patterns of the responses of the two models over time $t$, denoted as $C(t)$ (see Eq.~\eqref{SI:eq:cosinedissimilarity} in Sec.~\nameref{SI:ResultsDetails} for mathematical formulation).
A plot of $C(t)$ in Fig.~\ref{fig:timescales}B shows that the FNP maximally perturbs the dynamics at time $t \approx \SI{8}{\ms}$, which closely corresponds to the time at which the two waves from $\mathbf{p}$ and $\mathbf{q}$ meet.
We thus find that a single FNP has a strongly timescale-dependent perturbation on the model's dynamics, being substantial over short, millisecond timescales ($\lessapprox \SI{30}{\ms}$) but minimal on longer timescales ($> \SI{30}{\ms}$).

To understand why the stimulus-evoked dynamics are most strongly influenced by a stimulus-proximate FNP over short, millisecond timescales ($\lessapprox \SI{30}{\ms}$), we observed the evoked response of the geometric and hybrid models at two specific points $\mathbf{r}_1$ (in Fig.~\ref{fig:timescales}C(i)) and $\mathbf{r}_2$ (in Fig.~\ref{fig:timescales}C(ii)), which are positioned symmetrically on the interval from $\mathbf{p}$ to $\mathbf{q}$.
At both $\mathbf{r}_1$ and $\mathbf{r}_2$, the FNP's perturbation on the response commences at time $t \approx \SI{4}{\ms}$ (illustrated by the solid red line), through a reduction in activity at $\mathbf{r}_1$ and an increase in activity at $\mathbf{r}_2$.
The increase in activity at $\mathbf{r}_2$ arises from the additional traveling wave emanating from $\mathbf{q}$, which arrives at $\mathbf{r}_2$ before the wave emanating from $\mathbf{p}$.
Contrastingly, the decrease in activity at $\mathbf{r}_1$ arises from the total activity preservation property of the model, where the non-local propagation of activity from $\mathbf{p}$ to $\mathbf{q}$ reduces the amplitude of the traveling wave emanating from $\mathbf{p}$.
However, the perturbation induced by the FNP is mostly `canceled' once the waves emanating from $\mathbf{p}$ and $\mathbf{q}$ begin to interfere.
For $\mathbf{r}_1$ and $\mathbf{r}_2$, this cancellation occurs at $t \approx \SI{11}{\ms}$ (illustrated by the dotted red line in Figs~\ref{fig:timescales}C(i) and (ii)), where the negative perturbation at $\mathbf{r}_1$ is compensated for by the superposition of activity arriving from the secondary wave emanating from $\mathbf{q}$, and the positive perturbation at $\mathbf{r}_2$ is canceled by a weaker incident primary traveling wave from $\mathbf{p}$ from $t \approx \SI{11}{\ms}$.
This experiment indicates that, at any given point $\mathbf{r}$ on $\Omega$, the FNP can only substantially perturb the stimulus-evoked response over a specific time interval: from the time at which the first wavefront (from either $\mathbf{p}$ or $\mathbf{q}$) arrives at the point, to the time at which the second wavefront (from either $\mathbf{q}$ or $\mathbf{p}$, respectively) arrives.
By generalizing this result across all points on $\Omega$, we therefore expect that the perturbation $C(t)$ in Fig.~\ref{fig:timescales}B would be maximal at a time soon after $t = (L\sqrt{2}/8) / (r\gamma) \approx \SI{7}{\ms}$, when the two waves in the hybrid dynamics superimpose; but would become minimal from time $(L\sqrt{2} / 2) / (r\gamma) \approx \SI{28}{\ms}$, when the traveling wave from $\mathbf{p}$ reaches every point on $\Omega$.
These simple calculations capture the simulation results of Fig.~\ref{fig:timescales}B well and explain why an FNP strongly affects spatial activity patterns only over a relatively short time interval on short (millisecond) timescales.

\begin{figure}[!ht]
    \centering
    {\includegraphics[width = \textwidth]{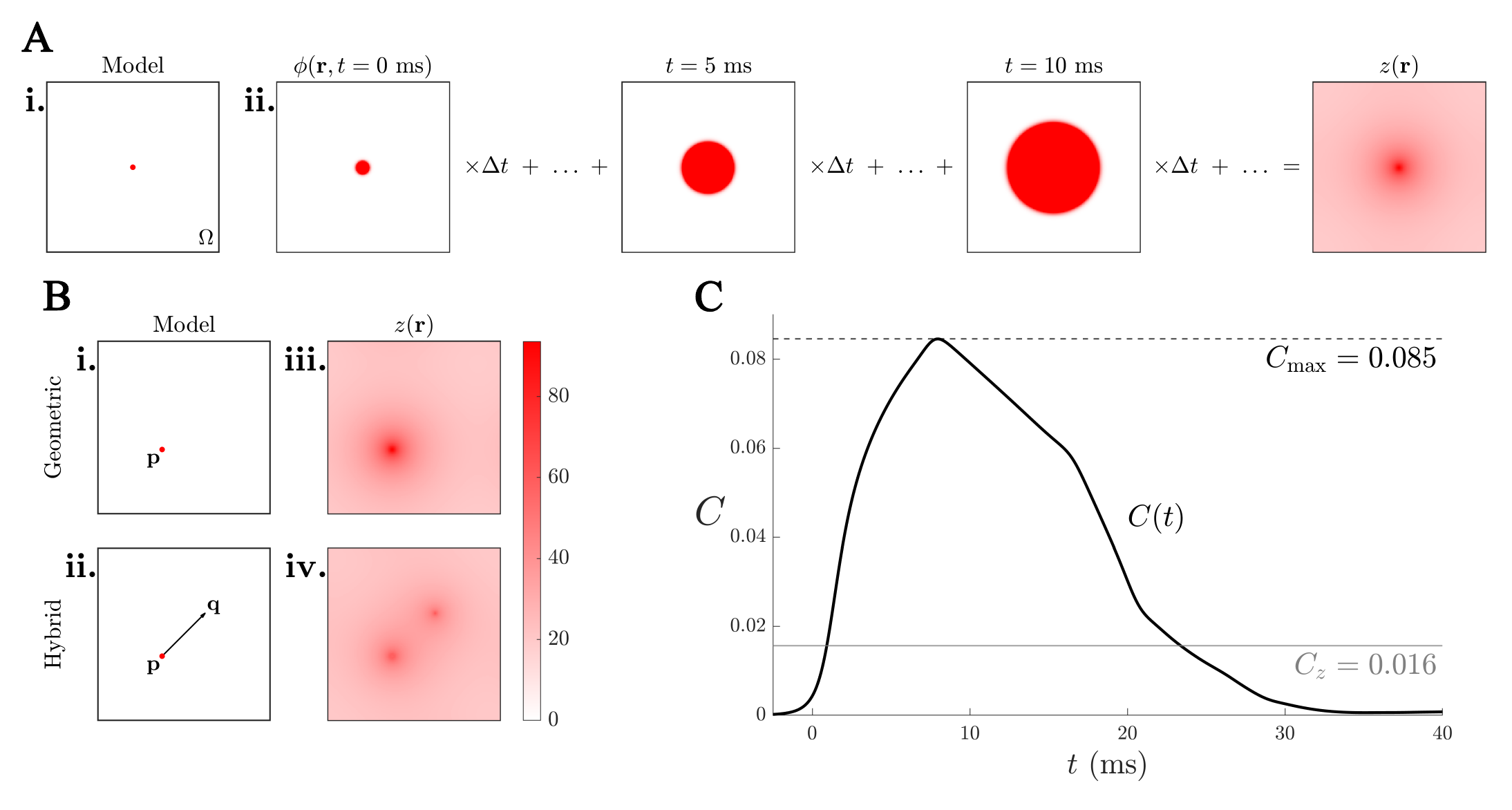}}\qquad
    \captionsetup{font=small}
    \caption{
    \textbf{When neural dynamics are resolved only over long timescales (such as those accessible using fMRI), a single FNP has a much smaller impact on measured neural dynamics.}
    \textbf{A.} The modeled BOLD response (BOLD signal observation of the stimulus-evoked response), denoted $z$, is computed as the zero-frequency temporal Fourier component of the model's stimulus-evoked response.
    \textbf{i.} To demonstrate the contrast between the stimulus-evoked response and the BOLD response, we apply an impulse stimulus (positioned at the red dot) to the geometric model.
    \textbf{ii.} The stimulus-evoked response of the geometric model, $\phi(\mathbf{r}, t)$, is then integrated over time to approximate the slow BOLD response, shown by the heatmap of $z(\mathbf{r})$.
    \textbf{B.} We modeled the evoked BOLD response of \textbf{i.} the geometric model, and \textbf{ii.} the hybrid model containing the FNP $\mathbf{p} \rightarrow \mathbf{q}$, when the stimulus is positioned at $\mathbf{p}$ (indicated by a red dot).
    \textbf{iii.} and \textbf{iv.} plot the BOLD response $z(\mathbf{r})$ as a heatmap for the geometric and hybrid models, respectively.
    \textbf{C.} The cosine dissimilarity between the BOLD responses of the geometric and hybrid models, denoted $C_z$ (Eq.~\eqref{SI:eq:cosinedissimilaritybold} in Sec.~\nameref{SI:ResultsDetails}), quantifies the perturbation of the FNP on the evoked BOLD response.
    $C_z$ is illustrated as the horizontal gray line, $C(t)$ from Fig.~\ref{fig:timescales}B is illustrated as the black curve, and $C_{\max} = \max_t\{C(t)\}$ is illustrated as the dotted black line.
    }
    \label{fig:fMRI}
\end{figure}

The findings above suggest that the perturbation of the FNP $\mathbf{p} \to \mathbf{q}$ on the model's stimulus-evoked dynamics would reduce in magnitude as the resolved dynamics becomes increasingly restricted to longer timescales, such as the order-of-seconds timescales resolved by fMRI \cite{robinsonDeterminationEffectiveBrain2014}.
We explicitly tested this prediction by examining the perturbation of the FNP on the model's BOLD response to the impulse stimulus, i.e., the BOLD signal observation of the stimulus-evoked response.
At each $\mathbf{r}$, the BOLD response was approximated as $z(\mathbf{r})$, the stimulus-evoked response's zero temporal frequency component \cite{robinsonDeterminationEffectiveBrain2014}, or equivalently, the time-integrated response as depicted schematically in Fig.~\ref{fig:fMRI}A (see Eq.~\eqref{SI:eq:bold} in Sec.~\nameref{SI:NumericalTreatment} for details on how $z(\mathbf{r})$ is computed numerically).
This linear approximation closely aligns with, and hence can be generalized to BOLD responses computed with more complex nonlinear hemodynamic forward models \cite{robinsonBOLDResponsesStimuli2006}.

Figures~\ref{fig:fMRI}B(i) and (ii) illustrate again the geometric and hybrid model respectively from Figs~\ref{fig:timescales}A(i) and (ii), and the impulse stimulus positioned at $\mathbf{p}$.
Figures~\ref{fig:fMRI}B(iii) and (iv) are heatmaps of the evoked BOLD response of both models to this stimulus, i.e., the time-integrated spatial activity pattern of the stimulus-evoked responses in Figs~\ref{fig:timescales}A(iii) and (iv).
The FNP's perturbation on the modeled BOLD response is concentrated near $\mathbf{p}$ and $\mathbf{q}$, increasing the response near $\mathbf{p}$ and decreasing it near $\mathbf{q}$, which is consistent with the same FNP's perturbation on the model's stimulus-evoked dynamics on longer timescales ($> \SI{30}{\ms}$) in our experiments above (Figs~\ref{fig:timescales}A(iii) and (iv)).
To quantify this perturbation, we measured $C_z$, the cosine dissimilarity between the BOLD responses of the geometric and hybrid model (see Eq.~\eqref{SI:eq:cosinedissimilaritybold} in Sec.~\nameref{SI:ResultsDetails} for mathematical detail).
A comparison between $C_z$ and $C(t)$ in Fig.~\ref{fig:fMRI}C shows that $C_z = 0.016$ is larger than the limiting value of $C(t)$ as $t \to \infty$, meaning that the FNP's strong perturbations over short timescales ($\lessapprox \SI{30}{\ms}$) are partially captured by the BOLD response.
However, $C_z$ is less than a fifth of $C_{\max} = \max_t \{C(t)\} = 0.085$, the maximum attained value of $C(t)$ at time $t \approx \SI{8}{\ms}$, indicating the short-timescale perturbation captured by the BOLD response is still substantially lower than what would be measured if the dynamics were properly time-resolved.
Hence, restricting our observations of the activity dynamics to the longer, order-of-seconds timescales captured by fMRI indeed limits our ability to access the shorter timescales over which FNPs most prominently shape the stimulus-evoked dynamics.
These findings provide a potential explanation for the discrepant empirical findings on the influence of FNPs on cortical dynamics across differing timescales: An FNP facilitates rapid non-local activity propagation on the sub-second timescales resolved by VSD imaging, EEG and calcium imaging \cite{ferezouSpatiotemporalDynamicsCortical2007, mohajeraniSpontaneousCorticalActivity2013, limVivoLargeScaleCortical2012, massiminiCorticalMechanismsLoss2012, seguinCommunicationDynamicsHuman2023, liuCellClassspecificLongrange2024}, but has a weaker influence on the longer timescales captured by fMRI, in which the idealized geometric connectivity rule of neural field models becomes a more valid approximation \cite{hendersonEmpiricalEstimationEigenmodes2022, pangGeometricConstraintsHuman2023}.

\subsection*{Influence of a single FNP on cortical dynamics driven by inputs of varying spatial precision}

We next investigated how the perturbation of a single FNP on the model dynamics depends on the spatial precision of the driving input.
We particularly explored perturbations on noise-driven dynamics (dynamics driven by noisy or temporally unstructured input), so that we could directly measure the FNP-induced perturbations to spontaneous activity, which is a commonly studied form of noise-driven dynamics where the input is spatially uniform \cite{foxSpontaneousFluctuationsBrain2007, uddinBringNoiseReconceptualizing2020}.

The noise-driven linear spatiotemporal dynamics of cortical activity can be decomposed as a superposition of linear evoked responses to spatially uncorrelated impulse stimuli across the cortical surface \cite{robinsonDeterminationEffectiveBrain2014, robinsonInterrelatingAnatomicalEffective2012, robinsonPhysicalBrainConnectomics2019}.
Hence, the perturbation of an FNP on noise-driven dynamics can be interpreted as its perturbation on stimulus-evoked dynamics, aggregated across all stimulus positions on the cortical surface, and weighted in accordance to the variance of the noisy input at each position.
The previous experiments in Fig.~\ref{fig:timescales} characterized the perturbation to geometric stimulus-evoked dynamics caused by a single FNP when the stimulus is incident at the FNP source location.
However, a sweep of all possible stimulus positions on $\Omega$ shows that the extent of this perturbation (quantified by both $C_{\max} = \max_t \{C(t)\}$ and $C_z$) decreases with its distance from the source of the FNP (Fig.~\ref{SI:fig:stim_position}), indicating that the FNP's non-local propagation is highly specific to activity that is proximate to its source location.
Under our aforementioned interpretation, this result in the stimulus-evoked setting therefore suggests that a single FNP's perturbation to noise-driven dynamics would depend not only on the configuration of the FNPs, but also on the spatial precision of the driving noisy input.
In particular, we predict that the FNP's perturbation on noise-driven dynamics increases as the input becomes more spatially localized at the FNP source, and decreases as the input becomes more spatially uniform in variance per spontaneous dynamics.

\begin{figure}[!ht]
    \centering
    {\includegraphics[width = \textwidth]{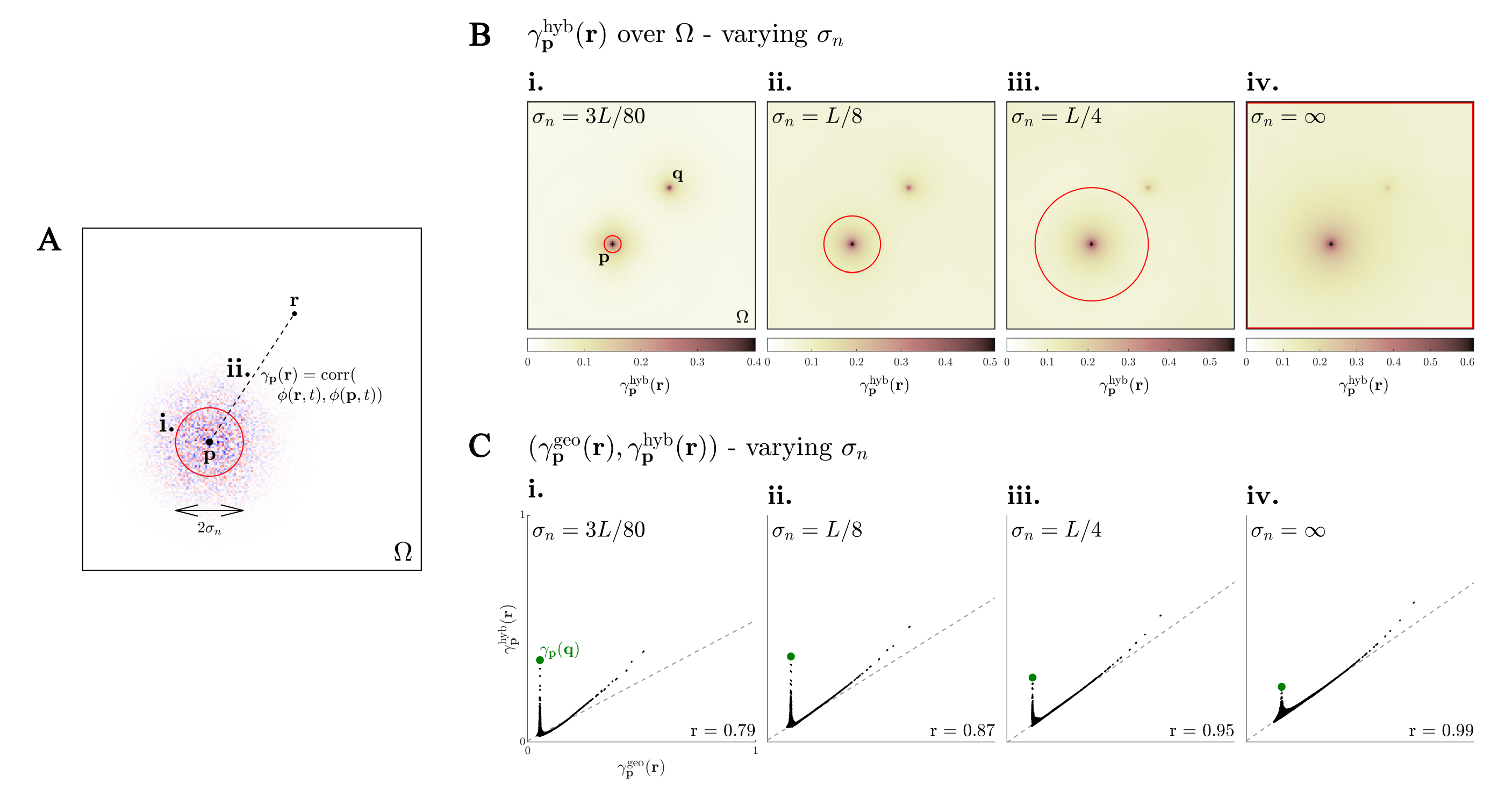}}\qquad
    \captionsetup{font=small}
    \caption{
    \textbf{The influence of an FNP on noise-driven dynamics is strongest for spatially precise input drives centered on the FNP source.}
    \textbf{A.} We investigated the influence of a single FNP $\mathbf{p} \to \mathbf{q}$ on noise-driven dynamics for noisy inputs for varying levels of spatial precision.
    \textbf{i.} We formulated the noisy input in Eq.~\eqref{eq:whitenoise} by setting $g(\mathbf{r}) = \exp (-\lVert \mathbf{r} - \mathbf{p} \rVert ^2 / 2\sigma_n^2)$, a Gaussian function with center $\mathbf{p}$ and spatial width $\sigma_n$ (indicated by the radius of the red circle).
    \textbf{ii.} We summarized the spatial correlation structure of the noise-driven model dynamics via the spatial correlation function to the reference point $\mathbf{p}$ (with other points $\mathbf{r}$), which we denote as $\gamma_\mathbf{p}(\mathbf{r})$.
    \textbf{B}: A heatmap of $\gamma_\mathbf{p}^\text{hyb}(\mathbf{r})$, the spatial correlation function with reference point $\mathbf{p}$ of the hybrid model containing the $\mathbf{p} \to \mathbf{q}$ FNP, for four different spatial widths of noisy input:
    \textbf{i.} $\sigma_n = 3L/80$;
    \textbf{ii.} $\sigma_n = L/8$;
    \textbf{iii.} $\sigma_n = L/4$; and
    \textbf{iv.} spatially uniform input ($\sigma_n \rightarrow \infty$).
    The $\sigma_n$ values are indicated by the radius of an annotated red circle for i., ii., and iii.
    For each heatmap, the maximum value of the color scale is the maximum attained value of $\gamma_\mathbf{p}^\text{hyb}(\mathbf{r})$.
    \textbf{C}: Scatter plots of the elements of the functions $\gamma_\mathbf{p}^\text{hyb}(\mathbf{r})$ and $\gamma_\mathbf{p}^\text{geo}(\mathbf{r})$ (the spatial correlation function with reference point $\mathbf{p}$ of the geometric model) for the four $\sigma_n$ values.
    In all scatter plots, the green dot corresponds to the point $(\gamma_\mathbf{p}^\text{geo}(\mathbf{q}),\gamma_\mathbf{p}^\text{hyb}(\mathbf{q}))$, the gray line is the linear fit between the elements, and $\mathrm{r}$ is the Pearson correlation.
    }
    \label{fig:spontaneous}
\end{figure}

To test and quantify this prediction, we investigated how the FNP $\mathbf{p} \to \mathbf{q}$ perturbs the dynamics of a range of noise-driven activity---from activity driven by noisy input spatially localized at $\mathbf{p}$, to spontaneous activity (activity driven by spatially uniform noisy input)---and compared the magnitude of the perturbation of the FNP across this variation of input precision.
As illustrated in Fig.~\ref{fig:spontaneous}A, we formulated the noisy input $f^\text{stim}$ in Eq.~\eqref{eq:whitenoise} by setting the  $g(\mathbf{r})$, the input variance, as a Gaussian with fixed center $\mathbf{p}$ (Fig.~\ref{fig:spontaneous}A(i)).
By varying the spatial width $\sigma_n$, we could assign different spatial profiles to the variance of the noisy inputs, ranging from spatially localized input at $\mathbf{p}$ ($\sigma_n \to 0$), to spatially uniform input (uniformly distributed in the limit $\sigma_n \to \infty$).
For any choice of $\sigma_n$, we then summarized the statistics of the resultant noise-driven activity via the spatial correlation function from reference point $\mathbf{p}$, denoted $\gamma_\mathbf{p}(\mathbf{r})$ (Fig.~\ref{fig:spontaneous}A(ii)), which measures the Pearson correlation between the activities at $\mathbf{p}$ and $\mathbf{r}$ for each $\mathbf{r} \in \Omega$ (see Sec.~\nameref{SI:ResultsDetails} for details on computation).
The reference point $\mathbf{p}$ was natural to focus on, as it allows us to understand how the FNP is perturbing the dynamics by propagating activity from $\mathbf{p}$ (to $\mathbf{q}$).

Figure~\ref{fig:spontaneous}B plots heatmaps of $\gamma_\mathbf{p}^\text{hyb}(\mathbf{r})$, the spatial correlation function of the hybrid model, for $\sigma_n = 3L/80, L/8, L/4$, and $\sigma_n \rightarrow \infty$ (corresponding to space-time white noise).
We first note for all $\sigma_n$ the expected decay of spatial correlation as a function of distance from $\mathbf{p}$, mediated by local geometric propagations of activity.
The FNP also expectedly perturbs the geometry-mediated noise-driven dynamics by mediating long-range correlations between $\mathbf{p}$ and $\mathbf{q}$, the FNP target, due to non-local propagation between the two points.
To quantify this perturbation, we compared $\gamma_\mathbf{p}^\text{hyb}(\mathbf{r})$ with $\gamma_\mathbf{p}^\text{geo}(\mathbf{r})$, the spatial correlation function of the geometric model with reference point $\mathbf{p}$, for the same selections of $\sigma_n$ (shown in Fig.~\ref{SI:fig:spontaneous_homogeneous}).
Comparisons of the elements of $\gamma_\mathbf{p}^\text{hyb}(\mathbf{r})$ and $\gamma_\mathbf{p}^\text{geo}(\mathbf{r})$ for each $\sigma_n$ are plotted as scatters in Figs~\ref{fig:spontaneous}C(i)--(iv) for each $\sigma_n$ value.
These plots show that the FNP's perturbation on noise-driven dynamics is observable in all cases, with the largest deviation in the linear fit between $\gamma_\mathbf{p}^\text{hyb}(\mathbf{r})$ and $\gamma_\mathbf{p}^\text{geo}(\mathbf{r})$ occurring at $\mathbf{q}$, denoted by the green dot.
However, the Pearson correlation between the elements of $\gamma_\mathbf{p}^\text{hyb}(\mathbf{r})$ and $\gamma_\mathbf{p}^\text{geo}(\mathbf{r})$, denoted $\mathrm{r}$, increases with $\sigma_n$.
The noise-driven dynamics of the geometric and hybrid models therefore becomes increasingly similar in correlation structure as the noisy input becomes more spatially uniform, characteristic of spontaneous activity.
This finding is in line with our expectation and suggests that differences in the spatial precisions of input may also contribute to conflicting empirical findings about the influence of FNPs on cortical dynamics: while an FNP exerts its strongest influence when the input is spatially localized at its source \cite{ferezouSpatiotemporalDynamicsCortical2007, mohajeraniSpontaneousCorticalActivity2013, limVivoLargeScaleCortical2012, liuCellClassspecificLongrange2024}, it has a weaker influence on spontaneous dynamics driven by spatially uniform inputs, which can be captured by a simpler geometric connectivity rule \cite{robinsonDeterminationDynamicBrain2021, hendersonEmpiricalEstimationEigenmodes2022, pangGeometricConstraintsHuman2023}.

\subsection*{Influence of a complex network of FNPs on cortical dynamics}

The results above demonstrate a single added FNP perturbs the model dynamics most strongly when the dynamics are observed on relatively short timescales ($\lessapprox \SI{30}{\ms}$) and when the driving input is spatially localized near the source of the FNP.
We now extend to a more realistic setting in which a connectome of multiple FNPs are distributed across the cortical sheet, facilitating rapid and non-local interactions across a complex distributed network.
This setting allows us to explore specific network properties of FNPs that violate the spatially homogeneous and isotropic connectivity rule commonly assumed by neural field models \cite{markovRoleLongrangeConnections2013, betzelSpecificityRobustnessLongdistance2018, vandenheuvelRichClubOrganizationHuman2011, hendersonUsingGeometryUncover2013}.

The perturbation of a single FNP on stimulus-evoked dynamics being strongest over timescales $\lessapprox \SI{30}{\ms}$ (in Fig.~\ref{fig:timescales}) could be explained by a reduction in perturbation over longer timescales ($> \SI{30}{\ms}$) following the superposition of traveling waves from the source and target of the FNP.
However, when multiple FNPs are present, the superposition of waves from the source and target of one FNP can be followed by a subsequent perturbation by an FNP at another location, as demonstrated in Fig.~\ref{fig:proofofprinciple} (using an example connectome of $50$ random FNPs).
Motivated by this possibility, here we aimed to quantitatively test whether the timescale-dependent perturbation of a single FNP, as demonstrated above, can be extended to a connectome of multiple FNPs.

\begin{figure}[!ht]
    \centering
    {\includegraphics[width = \textwidth]{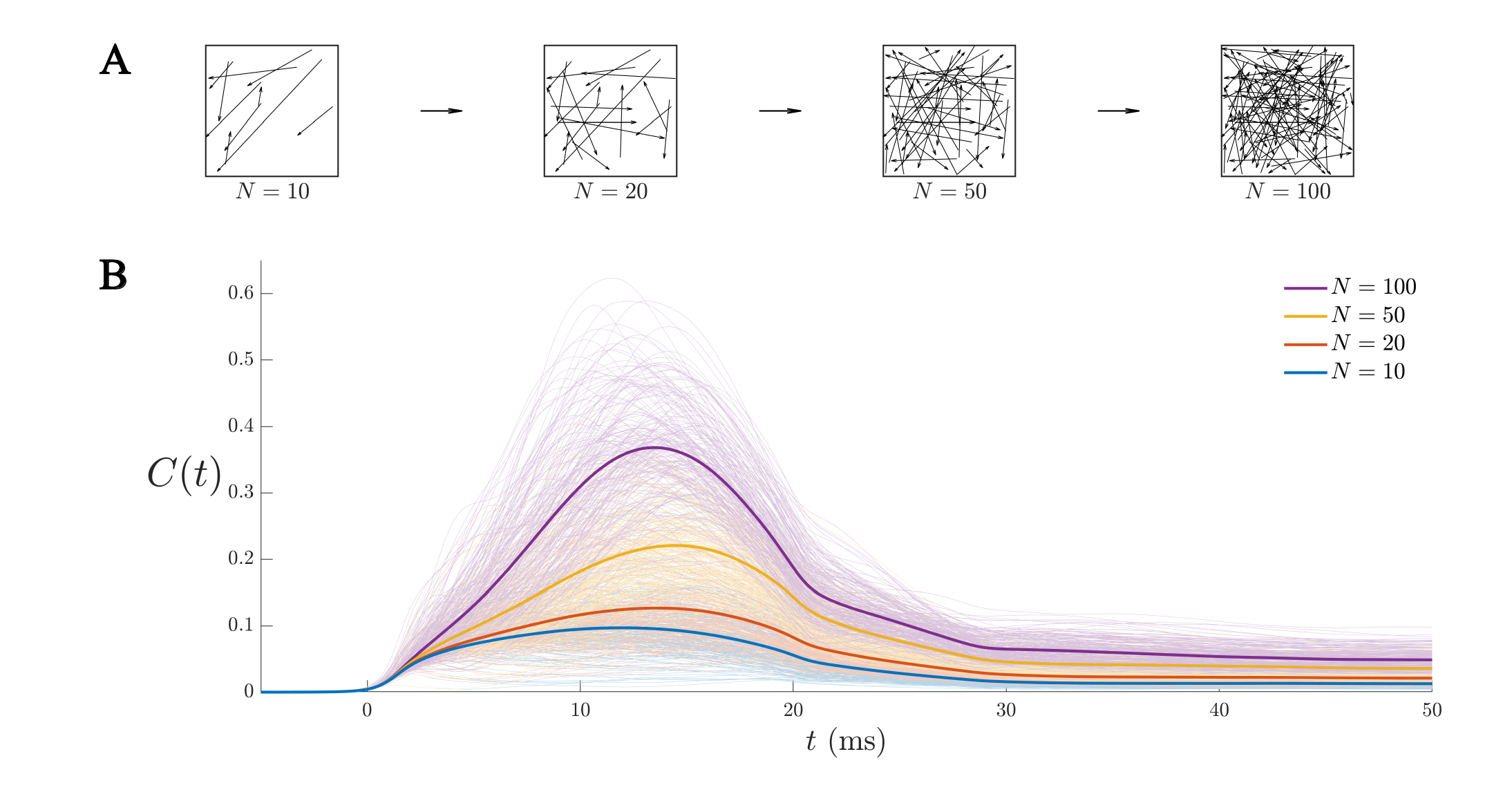}}\qquad
    \captionsetup{font=small}
    \caption{
    \textbf{A connectome of multiple randomly positioned FNPs perturbs geometric dynamics more strongly with the number of FNPs $N$ over timescales $\lessapprox \SI{30}{\ms}$.}
    \textbf{A.} For each of four different numbers of FNPs: $N = 10, 20, 50, 100$, we generated an ensemble of $200$ connectomes, with source and target locations sampled uniformly at random from $\Omega$.
    \textbf{B.} Realizations of $C(t)$ for each connectome in each of the four ensembles, with each ensemble's average shown as a thicker curve (computed as the ensemble average of $C(t)$ for a fixed $t$).
    }
    \label{fig:connectome_random}
\end{figure}

For a fixed number of FNPs $N$, we generated an ensemble of $200$ connectomes, each with $N$ randomly positioned FNPs with source and target positions sampled uniformly on $\Omega$.
An example of a generated connectome for each of $N = 10, 20, 50, 100$ FNPs is illustrated in Fig.~\ref{fig:connectome_random}A.
As above, we measured the perturbation of each connectome on the stimulus-evoked dynamics using $C(t)$ (cf. Eq.~\eqref{SI:eq:cosinedissimilarity}, Sec.~\nameref{SI:ResultsDetails}), in a stimulus-response setting using an impulse stimulus positioned at the source location of the first randomly sampled FNP.
This stimulus position was selected to ensure that the perturbation of all sampled connectomes begins from the same time (the time of stimulus onset), so that the $C(t)$ curves align and can be compared straightforwardly.

The perturbation curves $C(t)$ across ensembles of connectomes for each of $N = 10, 20, 50, 100$ are shown in Fig.~\ref{fig:connectome_random}B.
For all $t$, we find that $C(t)$ tends to increase with the number of FNPs $N$.
This finding is consistent with our expectation, since increasing $N$ increases the expected number of rapid non-local activations during the response.
$C(t)$ for all $N$ also tends to increase to a maximum value at a time between $\SI{10}{\ms}$ and $\SI{20}{\ms}$.
Note that this time to maximum is larger than that attained with a single FNP ($\approx \SI{8}{\ms}$, see Fig.~\ref{fig:timescales}B), reflecting the different times from stimulus onset at which each FNP perturbs the dynamics.
However, after reaching the maximum, $C(t)$ for all $N$ tends to reduce to a smaller limiting value over longer timescales ($> \SI{30}{\ms}$).
This subsequent behavior of $C(t)$ for each $N$ therefore demonstrates the timescale-dependent perturbation of a connectome of multiple randomly positioned FNPs on cortical dynamics, with the most salient perturbation occurring over shorter timescales ($\lessapprox \SI{30}{\ms})$.

In our investigations of noise-driven dynamics (in Fig.~\ref{fig:spontaneous}), a single FNP was shown to contribute a relatively small perturbation to geometrically mediated spontaneous dynamics (driven by a noisy input of spatially uniform variance).
This was explained by the inverse relationship between the perturbation ($C_{\max}$ or $C_z$) and the distance between the stimulus position and the source of the FNP (Fig.~\ref{SI:fig:stim_position}).
However, connectomic data indicates that FNPs tend to organize into non-random topological configurations, leading to spatially inhomogeneous and anisotropic connectivity patterns that violate the average connectivity rule typically used in neural field models \cite{hendersonUsingGeometryUncover2013}.
A well-described characteristic of cortical structural connectivity is the non-uniform distribution of FNPs, which tends to concentrate connectivity on a subset of highly connected cortical regions called `hubs' \cite{markovRoleLongrangeConnections2013, betzelSpecificityRobustnessLongdistance2018, vandenheuvelRichClubOrganizationHuman2011, arnatkeviciuteGeneticInfluencesHub2021}.
Hub--hub connectivity makes the existence of chains of inter-linked FNPs (i.e., where the target of one FNP is proximate to the source of another FNP) more likely and thus facilitating distributed, multi-FNP interactions.
Such effects would contribute differently in hub areas (where multi-FNP chains are more likely than in random networks) than in non-hub areas (where multi-FNP chains are less likely than in random networks).
Accordingly, we wanted to investigate how the concentration of connectivity on specific spatial locations may lead to higher-order network effects (beyond that of a single FNP) that result in different behavior to that observed for single FNPs in isolation.

Relative to random (unconstrained) connectivity, we explored the influence on spontaneous dynamics (quantified as $C_{\max}$) of three key connectomic constraints: the exponential distance rule (indexed by parameter $\lambda_e$), hub specificity (indexed by parameter $\lambda_h$), and rich-club specificity (indexed by parameter $\lambda_r$).
For a given $\lambda_e, \lambda_h$, or $\lambda_r$, connectomes were generated with a rejection sampling algorithm (see Sec.~\nameref{SI:ResultsDetails} for details on each algorithm).
We explain each constraint in turn.

\begin{figure}[!ht]
    \centering
    {\includegraphics[width = \textwidth]{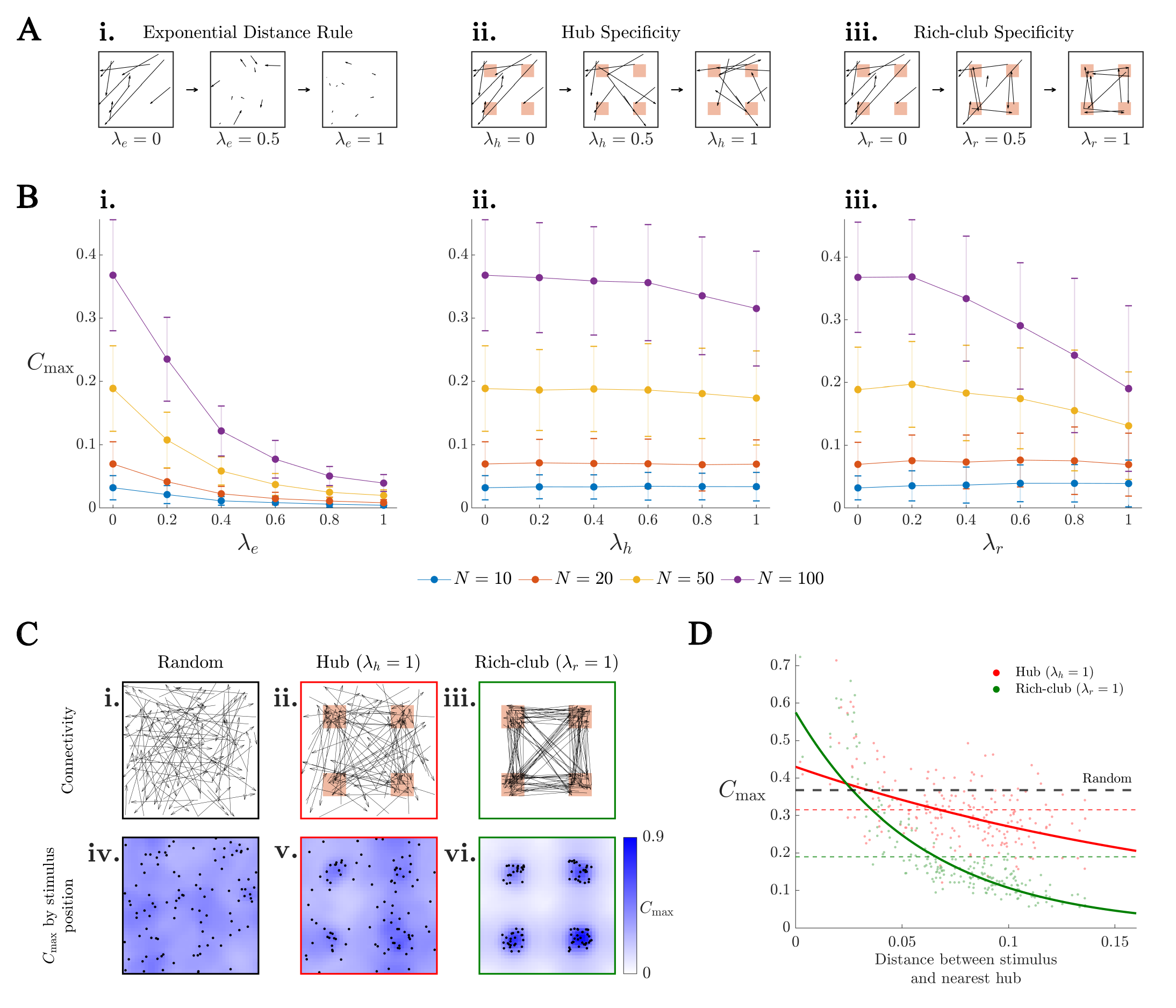}}\qquad
    \captionsetup{font=small}
    \caption{
    \textbf{The influence of a connectome of FNPs on spontaneous dynamics is either unaffected or reduced by the incorporation of distance-dependent or positional specificity-based connectomic constraints}.
    \textbf{A.} We examine the contribution of three connectivity constraints using a parameter that quantify the extent of the constraint: $\lambda_e, \lambda_h, \lambda_r$.
    \textbf{i.} An exponential distance rule (EDR) constraint is captured by the parameter $\lambda_e$, which controls the rate at which connection probability decays with separation distance.
    \textbf{ii.} A hub specificity constraint is captured by the parameter $\lambda_h$, which controls the probability that a given FNP is incident with a hub area (colored orange).
    \textbf{iii.} A rich-club specificity constraint is captured by the parameter $\lambda_r$, which controls the probability that a given FNP connects a pair of hubs.
    \textbf{B.} The distribution of $C_{\max}$ over an ensemble of $200$ connectomes containing $N$ FNPs each and $200$ uniformly sampled stimulus positions, for increasing values of: \textbf{i.} $\lambda_e$;
    \textbf{ii.} $\lambda_h$; and
    \textbf{iii.} $\lambda_r$.
    Each distribution is represented with error bars of one standard deviation from the mean.
    \textbf{C.} We compared the distribution of $C_{\max}$ for different stimulus positions, across three sampled connectomes:
    \textbf{i.} A random (unconstrained) connectome; 
    \textbf{ii.} A maximally hub constrained connectome ($\lambda_h = 1$); and
    \textbf{iii.} A maximally rich-club constrained connectome ($\lambda_r = 1$); each containing $N = 100$ FNPs
    \textbf{iv.}, \textbf{v.}, \textbf{vi.} Heatmaps of $C_{\max}$ by stimulus position of the sampled random, hub-constrained, and rich-club-constrained connectomes shown in i., ii., and iii., respectively.
    The stimulus positions were a $40 \times 40$ grid of equally spaced points on $\Omega$, $\{(i L/40, j L/40):i, j = 1, \ldots, 40\}$.
    Source positions of all FNPs are indicated by black dots.
    \textbf{D.} Scatter plots of the sampled values of $C_{\max}$ and the distance between the sampled stimulus and center of the nearest hub, for an ensemble of $200$ maximally hub constrained ($\lambda_h = 1$) connectomes and $200$ maximally rich-club constrained ($\lambda_r = 1$) connectomes.
    The mean of both distributions is annotated using a dotted horizontal line, and the average value of $C_{\max}$ over an ensemble of $200$ random connectomes (with no hubs).
    Exponential fits to the data points from the maximally hub and rich-club constrained ensemble are annotated by the solid curves.
    }
    \label{fig:connectome_nonrandom}
\end{figure}

The exponential distance rule (EDR) captures the empirical finding that the probability of a FNP existing between two given cortical populations decays exponentially as a function of the length of the projection \cite{ercsey-ravaszPredictiveNetworkModel2013, horvatSpatialEmbeddingWiring2016, fulcherTranscriptionalSignatureHub2016}.
To generate a connectome with a given EDR decay rate, we used a decay parameter $\lambda_e$ to control the rate at which the probability of an FNP existing between two given points on $\Omega$ decays exponentially with its (Euclidean) separation distance (accounting for periodic boundary conditions) (see \eqref{SI:eq:edr} in Sec.~\nameref{SI:ResultsDetails} for formulation of this probability).
Examples of three different connectomes with $10$ FNPs and varying EDR decay rates, generated from $\lambda_e = 0, 0.5, 1$, are shown in Fig.~\ref{fig:connectome_nonrandom}A(i), demonstrating the increasing penalization of long-range connections with increasing $\lambda_e$.

Hub specificity captures the empirical observation that specific hubs of the cortex tend to be particularly strongly connected and are thought to play a major role in mediating distributed information transfer across the cortex \cite{markovRoleLongrangeConnections2013, betzelSpecificityRobustnessLongdistance2018, betzelGenerativeModelsHuman2016}.
To generate a connectome with a given level of hub specificity, we used a scaling parameter $\lambda_h$ to assign the probability that each FNP connects a population inside a hub to (or from) a population outside the same hub; ranging from the probability attained by chance from uniform sampling ($\lambda_h = 0$) to certainty ($\lambda_h = 1$) (see \eqref{SI:eq:hubspecificity} in Sec.~\nameref{SI:ResultsDetails} for formulation of this probability).
For simplicity, we modeled the case of four hub areas (which also accords with some early work on hub identification in the human cortex \cite{hagmannMappingStructuralCore2008, iturria-medinaStudyingHumanBrain2008}) which were chosen as square areas on the domain $\Omega$.
As depicted in Figs~\ref{fig:connectome_nonrandom}A(ii), hub areas were allocated with centers $(L/2,L/2) + (\pm L/4,\pm L/4)$ to maximize spacing between hubs, and side length $\sqrt{34}$ since the cortex of each cerebral hemisphere was parcellated into $34$ regions \cite{hagmannMappingStructuralCore2008, iturria-medinaStudyingHumanBrain2008}.
Examples of three different connectomes with $10$ FNPs for $\lambda_h = 0, 0.5, 1$ are illustrated in Fig.~\ref{fig:connectome_nonrandom}A(ii).
At $\lambda_h = 0$ we see spatially uniform random FNP connectivity unaffected by the assignment of hub areas, whereas at $\lambda_h = 1$ we see that all FNPs connect with a hub.

Rich-club specificity captures the empirical phenomenon of dense inter-hub connectivity, resulting in a rich-club core of the connectome \cite{vandenheuvelHighcostHighcapacityBackbone2012, vandenheuvelRichClubOrganizationHuman2011, arnatkeviciuteGeneticInfluencesHub2021}.
To generate a connectome with a given level of rich-club specificity, we used a parameter $\lambda_r$ to assign the probability that each FNP connects two distinct hubs; ranging from the probability attained by chance from spatially uniform random connectivity ($\lambda_r = 0$) to certainty ($\lambda_r = 1$) (see Eq.~\eqref{SI:eq:richclubspecificity} in Sec.~\nameref{SI:ResultsDetails} for formulation of this probability).
Example realizations of connectomes containing $10$ FNPs for $\lambda_r = 0, 0.5, 1$ are plotted in Fig.~\ref{fig:connectome_nonrandom}A(iii).

We investigated how incorporating each of these three connectomic constraints (EDR, $\lambda_e$; hub specificity, $\lambda_h$; or rich-club specificity, $\lambda_r$) affects the perturbation to the model's spontaneous dynamics.
We generated an ensemble of $200$ connectomes each containing $N = 10, 20, 50, 100$ FNPs, now repeating for different values of $\lambda_e \in [0, 1]$ (for the EDR constraint), $\lambda_h \in [0, 1]$ (for the hub specificity constraint), or $\lambda_r \in [0, 1]$ (for the rich-club specificity constraint).
We then quantified the average perturbation of each ensemble on the modeled spontaneous dynamics.
Specifically, we measured the average perturbation of each connectome in the ensemble on stimulus-evoked dynamics, for a uniformly sampled stimulus position on $\Omega$.
This quantification approach, which draws on the relationship between noise-driven activity and stimulus-evoked dynamics demonstrated in the previous subsection, is computationally faster than the conventional approach of computing correlation structures of spontaneous dynamics (see Fig.~\ref{SI:fig:connectome_nonrandom_corr} for similar results of the experiments using correlation structures for a smaller ensemble sample size).
Accordingly, for each sampled connectome we sampled an impulse stimulus uniformly at random from $\Omega$, and measured $C_{\max} = \max_t \{C(t)\}$ (cf. Eq.~\eqref{SI:eq:cosinedissimilarity}, Sec.~\nameref{SI:ResultsDetails}), the maximum perturbation that the connectome induces on the evoked response to this stimulus over time if the dynamics was resolved to sufficiently fine timescales.

Distributions of $C_{\max}$ as a function of $\lambda_e$, $\lambda_h$, and $\lambda_r$ are plotted in Figs~\ref{fig:connectome_nonrandom}B(i) and (ii).
For a fixed $\lambda_e, \lambda_h$, or $\lambda_r$, we see an increase in the average $C_{\max}$ with $N$ (the number of FNPs), matching the increase in perturbation with $N$ across timescales shown above for the random connectome (Fig.~\ref{fig:connectome_random}B).
The decreasing in $C_{\max}$ for increasing $\lambda_e$---i.e., as progressively shorter FNPs are favored---matches the expectation that fast-conducting perturbations induced by FNPs increase when the FNPs propagate activity over longer distances.
Indeed, in the limiting case of $\lambda_e \to \infty$, the average length of each sampled FNP approaches zero, and in turn $C_{\max}$ approaches zero.

Interestingly, the average $C_{\max}$ did not increase with $\lambda_h$, nor with $\lambda_r$, remaining relatively constant for small $N = 10, 20$, and decreasing for larger $N = 50, 100$ (Figs~\ref{fig:connectome_nonrandom}B(ii) and (iii)).
To better understand this result, we investigated how $\lambda_h$ and $\lambda_r$ affected the dependence of $C_{\max}$ on the position of the stimulus.
To this end, we sampled and studied three connectomes:
a random (unconstrained) connectome (Fig.~\ref{fig:connectome_nonrandom}C(i)); a maximally hub constrained connectome, $\lambda_h = 1$, in which every FNP connects a hub (Fig.~\ref{fig:connectome_nonrandom}C(ii)); and a maximally rich-club constrained connectome, $\lambda_r = 1$, in which every FNP connects a pair of hubs (Fig.~\ref{fig:connectome_nonrandom}C(iii)).
We constructed all three connectomes with $N = 100$ FNPs, since $C_{\max}$ was most sensitive to $\lambda_h$ and $\lambda_r$ in Figs~\ref{fig:connectome_nonrandom}B(ii) and (iii) for this value of $N$.
Figures~\ref{fig:connectome_nonrandom}C(iv)--(vi) show heatmaps of $C_{\max}$ by stimulus position for the random, hub-constrained ($\lambda_h = 1$), and rich-club constrained ($\lambda_r = 1$) connectomes, respectively.
Compared to random connectivity, the hub constrained and rich-club constrained connectomes exhibit a more spatially heterogeneous distribution of $C_{\max}$, with higher $C_{\max}$ values for stimuli targeted on or near hub areas, and lower $C_{\max}$ values as the stimulus becomes more distant from the hubs.
This spatial non-uniformity can be explained by the fact that a single FNP perturbs stimulus-evoked dynamics most strongly when the stimulus is positioned at the FNP source location (Fig.~\ref{SI:fig:stim_position}), with the non-random connectomes displaying a higher density of FNP sources (indicated by the black dots) at hub areas.
Figure~\ref{fig:connectome_nonrandom}D demonstrates more concretely that the spatial nonuniformity of $C_{\max}$ attained from hub specificity and rich-club specificity persists across an ensemble of $200$ maximally hub constrained connectomes ($\lambda_h = 1$) and $200$ maximally rich-club constrained ($\lambda_r = 1$) connectomes.
Consistent with the concentration of high $C_{\max}$ near hub locations, the plot explicitly demonstrates the decreasing relationship between $C_{\max}$ and the distance between the stimulus and the nearest hub for the hub constrained and rich-club constrained models (shown by solid curves).
However, when compared to a random ensemble of connectomes (dashed horizontal lines in Fig.~\ref{fig:connectome_nonrandom}D) the non-random connectomes tend to have larger values of $C_{\max}$ for stimulus positions proximate to a hub (stimulus--hub distances $<\SI{0.03}{\m}$), and smaller values of $C_{\max}$ for stimulus positions distant from a hub (stimulus--hub distances $>\SI{0.03}{\m}$).
Aggregating across all stimulus locations results in the hub and rich-club constrained ensembles having a lower average $C_{\max}$ than the random ensemble, indicating that the decrease in $C_{\max}$ for stimuli distant from hub locations outweighs the increase in $C_{\max}$ for stimuli proximate to hubs.

The results from Fig.~\ref{fig:connectome_nonrandom}C and D in the stimulus-evoked setting suggest that the hub and rich-club constraints result in a corresponding spatial specificity of the connectomic perturbation to spontaneous dynamics, which becomes concentrated on hubs.
However these spatially localized perturbations are largely `washed out' over larger lengthscales characteristic of the entire cortical surface, which is likely attributable to the fact that the hubs, where the perturbations increase, form a small proportion of the cortical surface ($4/34 \approx 12\%$ in our experiments).
Taken together, while the hub and rich-club constraints can amplify the perturbation of FNPs on evoked responses to (specifically) hub stimulation, it has a comparatively weaker effect on the perturbation to spontaneous dynamics.
This suggests that the impact of hub and rich-club connectivity on model dynamics depends on the spatial precision of the input driving the dynamics.

In summary, a connectome of multiple FNPs, that facilitates rapid and distributed communication across the cortex, perturbs the geometric dynamics more strongly with an increased number of FNPs but---as with prior results using a single FNP---these perturbations remain most prominent on timescales $\lessapprox \SI{30}{\ms}$.
Furthermore, relative to randomly embedded networks, simple models of spatially and topologically constrained connectomes (including hub and rich-club specificity) tend to concentrate the perturbative impact of FNPs on spontaneous dynamics at specific hub regions, but did not tend to elevate the aggregate perturbation over the entire cortical surface (and actually tended to reduce the spatially aggregated perturbation when the connectome contained sufficiently large numbers of FNPs, $N = 50, 100$).
Our findings provide a plausible and testable account for why the spatially homogeneous and isotropic average connectivity rule of neural field models can approximate resting-state fMRI dynamics (spontaneous dynamics over timescales $\gtrsim \SI{1}{\s}$), despite the presence of a spatially embedded and topologically complex network of FNPs.

\section{Discussion}

This work aims to address two lines of evidence regarding the functional role of fast-conducting non-local projections (FNPs) on cortical dynamics.
One range of experiments demonstrate the critical role of specific, genetically inherited FNPs in shaping the intricate spatiotemporal dynamics of cortical activity, such as the clear non-local activations seen in calcium-imaging of stimulus-evoked cortical responses in mouse \cite{ferezouSpatiotemporalDynamicsCortical2007, mohajeraniSpontaneousCorticalActivity2013,massiminiCorticalMechanismsLoss2012,limVivoLargeScaleCortical2012,seguinCommunicationDynamicsHuman2023,liuCellClassspecificLongrange2024}, or long-range correlations between cortical populations mediated by structural connectivity \cite{honeyPredictingHumanRestingstate2009, greiciusRestingStateFunctionalConnectivity2009, mohajeraniSpontaneousCorticalActivity2013}.
These empirical results, across scales and species, demonstrate the important role of FNP networks in facilitating the complex, spatially distributed cortico-cortical communication thought to underpin the brain's remarkable cognitive functionality, despite their costs to develop, maintain and operate \cite{vandenheuvelHighcostHighcapacityBackbone2012, vertesGeneTranscriptionProfiles2016, fulcherTranscriptionalSignatureHub2016, arnatkeviciuteGeneticInfluencesHub2021}.
On the other hand, a second collection of results demonstrates the success of neural field models which, via a spatially homogeneous and isotropic connectivity rule assumption \textit{neglect} specific FNPs, but can nevertheless predict many aspects of cortical dynamics accurately, particularly the spontaneous fluctuations observed on the order-of-seconds timescales of fMRI \cite{robinsonDeterminationDynamicBrain2021, hendersonEmpiricalEstimationEigenmodes2022, pangGeometricConstraintsHuman2023}.
This work sought to address the question: how can models of cortical dynamics that `smear out' the anatomical specificity of FNPs to a geometric average nevertheless generate such accurate predictions of cortical dynamics?

To this end, we introduced a new mathematical model of macroscale cortical activity (on millimeter lengthscales and above) which incorporates both continuous geometric propagation (as traveling waves along the cortical surface and discrete connectomic propagation (as rapid, non-local interactions between specific remote populations mediated by FNPs).
Our experiments demonstrate that the perturbations that FNPs induce on the model's geometrically-constrained cortical dynamics is both timescale dependent---being strongest on short timescales $\lessapprox \SI{30}{\ms}$---and dependent on the spatial precision of the input drive---being strongest for spatially localized inputs that are proximate to FNPs.
These settings in which FNPs strongly shape cortical dynamics---fast dynamical responses to spatially targeted inputs---suggest rapid non-local interactions as a core mechanism for facilitating the fast and spatially distributed information processing of specific sensory input drives (e.g., specific cortical inputs from `core' thalamic cells \cite{shineImpactHumanThalamus2023}).
Conversely, in settings such as resting-state fMRI, which measures spontaneous dynamics over slower order-of-seconds timescales, the cortical dynamics increasingly resemble that of geometric propagation (i.e., as if there were no FNPs).
Our model thus provides a novel and powerful way to characterize the effect of FNPs as specific spatial perturbations to geometric `mean field' dynamics; compared to models that do not include geometric dynamics and thus attribute all non-trivial dynamics to specific FNPs.
Our findings thus provide a plausible mechanistic explanation for why different experiments (on different spatial and temporal scales, and examining spontaneous versus stimulus-response dynamics) can yield different conclusions regarding the role of FNPs, which our experiments indicate is highly dependent on the timescale resolved by measurement and the spatial precision of the input drive.

Our results broadly match qualitative intuitions from the statistical mechanics of physical systems, which provides a way to understand how simple macroscopic behavior can emerge from complex microscopic interactions.
This includes the phenomenon of `universality', in which physical systems with different microscopic interactions nevertheless display common macroscale behavior (which can therefore exhibit surprising robustness to variations in microscopic physics) \cite{sethnaStatisticalMechanicsEntropy2021}.
A commonly studied example is the relatively simple Navier--Stokes equations, which emerge from extremely complex interactions between fluid molecules at the microscale, and which well-describe the macroscopic behavior of fluids regardless of the specific type of molecules (and their interactions).
In qualitative accordance with this loose expectation, here we find that as cortical inputs become less precise in space (as in spontaneous dynamics) and are less accurately resolved in time (as in modalities like fMRI), the specific microscopic wiring configurations of individual structural fibers of the cortex matter less than the spatially averaged rules that govern them.
Future work analyzing multiscale patterns in neural systems \cite{munnMultiscaleOrganizationNeuronal2024}, including relationships between scales, could investigate these scale-dependent relationships more directly and in more detail.

Our work suggests a potential reconciliation of conflicting perspectives on how best to build dynamical models of macroscale cortical activity constrained by underlying structural connectivity \cite{decoDynamicBrainSpiking2008, breakspearDynamicModelsLargescale2017, griffithsWholeBrainModellingPresent2022}.
On one hand, connectome-based neural mass models treat the cortex as a discrete network of lumped nodes, representing parcellated regions, connected by edges that accumulate all structural fibers connecting the corresponding regions \cite{honeyNetworkStructureCerebral2007, ghoshNoiseRestEnables2008, decoKeyRoleCoupling2009, abdelnourNetworkDiffusionAccurately2014, chaudhuriLargeScaleCircuitMechanism2015, decoDynamicsRestingFluctuations2017}.
This representation allows for explicit incorporation of specific structural connections from connectomic data, but artificially enforces a discretization (via a parcellation) with hard boundaries on a physically continuous object.
By contrast, neural field models treat the cortex as a two-dimensional continuum, and typically embed a geometric (spatially homogeneous and isotropic) structural connectivity onto this continuum \cite{nunezNeocorticalDynamicsHuman1995, robinsonPropagationStabilityWaves1997, jirsaSpatiotemporalForwardSolution2002}.
This representation retains the continuity of cortical tissue at the millimeter lengthscale without enforcing spatial discretization into parcels \cite{robinsonPhysicalBrainConnectomics2019}, but its idealized distance-dependent connectivity assumption cannot fully capture the specificity of structural connectivity, particularly that of FNP connectivity.
The findings of this work suggest that the continuous geometric approximation of structural connectivity (akin to neural field models) becomes a stronger approximation for capturing the spatiotemporal patterns of spontaneous cortical dynamics on longer timescales (like that of resting-state fMRI), while the additional non-local specificity of the connectome (which can be incorporated in neural mass models) becomes more crucial for capturing the cortex's more rapid processing of spatially localized stimuli.
Note that this does not imply that the specificity of the connectome is unimportant for capturing resting-state fMRI dynamics---indeed, the further constraining of geometric connectivity with specific connections from connectomic data can improve predictions of resting-state fMRI dynamics \cite{decoRareLongrangeCortical2021, vohryzekHumanBrainDynamics2025, decoNonlocalSchrodingerDiffusion2025}.
Rather, our findings highlight stimulus-evoked response dynamics as an experimentally challenging but important regime for developing and refining future dynamical models of macroscale brain activity — a regime in which the specificity of structural connectivity may play a more significant role than in resting-state fMRI \cite{maranAnalyzingBrainsDynamic2025}.

An important finding of this work is that a single FNP most strongly perturbed the model's geometric dynamics over short millisecond timescales ($< \SI{10}{\ms}$), and weakened in influence on longer timescales.
This finding may have direct implications for how FNPs perturb the spatial patterns of the geometric eigenmodes of cortical activity---the natural dynamical modes of activity derived from the cortex's geometric or spatial structure \cite{robinsonEigenmodesBrainActivity2016}.
If FNPs primarily influence the model's faster dynamics, we predict that the addition of FNPs (from connectomic data) to the cortex's geometric structure would induce perturbations to the geometric eigenmodes that increase with eigenmode order.
This prediction accords with the results of another mathematical investigation, which modeled the perturbation of a single FNP in a way that---like our model---preserved the spacetime-integrated propagated activity \cite{robinsonNearcriticalCorticothalamicEigenmodes2025}.
However, since the geometric eigenmodes of higher order contribute less power to spontaneous activity \cite{mullerMusicHemispheresCortical2022}, we suspect that the improvements that the connectome-perturbed eigenmodes provide in reconstructing measured activity would be dominated by physiological noise and measurement artifacts, both of which are major sources of error carried by resting-state fMRI measurements \cite{liuNoiseContributionsFMRI2016}.
If true, it would be a challenging feat to identify an alternative basis of connectome-perturbed eigenmodes that significantly improves the reconstruction accuracy of resting-state fMRI from simpler geometric eigenmodes, which is found to be the case in ongoing work 
\cite{vohryzekHumanBrainDynamics2025, sinamansourEigenmodesBrainRevisiting2024, faskowitzCommentaryPang20232023, patilCommentaryPang20232023}.
Nevertheless, it remains important for future work to investigate how the addition of FNPs derived from different structural brain properties actually perturbs the spatial patterns of geometric eigenmodes, as a means to mechanistically interpret how these perturbations impact the eigenmodes' reconstruction accuracy of measured activity.
 
This work also showed that a single FNP exerted its maximal perturbing effect on geometric dynamics when the input stimulus was localized precisely at the source location of the FNP.
This result potentially adds specificity to the prevailing hypothesis that the brain evolved costly FNPs between specific cortical populations to enhance computational efficiency beyond that achievable with purely geometric connectivity \cite{satoLongrangeConnectionsEnrich2021}; namely, that an FNP may play its most critical role in processing specific inputs arriving at or near its source.
Future work could investigate this hypothesis by investigating how the addition of a single FNP to geometric connectivity would affect the reconstructions of dynamic functional connectivity---functional connectivity over windows of time \cite{hutchisonDynamicFunctionalConnectivity2013}---during a sensory task consisting of a sequence of varying incoming sensory stimuli.
Our model predicts that the most pronounced deviations between the reconstruction accuracies of dynamic functional connectivity by the geometric and FNP-added eigenmodes would occur during the times when the stimulus is proximate to the source of the added FNP.

Our results also highlight how a fixed architecture of structural connectivity in the cortex is capable of supporting a vast array of spontaneous dynamical regimes, depending on how the system is driven.
This finding underscores an underappreciated specificity of the contribution of a given FNP in shaping cortical dynamics on the spatial properties of the cortical input drive (e.g., from the thalamus).
For the commonly studied case of spontaneous dynamics, typically modeled as spatially uniform cortical input \cite{sanz-leonNFTsimTheorySimulation2018, ghoshNoiseRestEnables2008}, the specific (anisotropic and inhomogenous) configurations of FNPs make surprisingly little effect on the resulting spontaneous dynamics, particularly those that can be resolved on the long timescales of fMRI.
However, our modeling highlights how the same, fixed underlying structural connectivity can support a much wider repertoire of spontaneous dynamical regimes, accessible by controlling the spatial properties of the input drive.
In this way, some FNPs can be relatively `silent' in shaping to the dynamics to some (distal) focal inputs, despite corresponding to major fiber tracts with `high weight' in a connectomic representation.
Our modeling thus suggests a more fluid picture of `the structure-function relationship' in which controllable inputs to the cortex can be used to facilitate a broad array of different types of distributed information processing that may be functionally suited to a given situation (e.g., balancing integration versus segregation on different timescales).
One possible way in which the brain may take advantage of this flexibility is by adjusting the spatial profiles of core (spatially precise) and matrix (spatially diffuse) thalamic inputs \cite{mullerDiffuseNeuralCoupling2020, mullerNonspecificMatrixThalamus2023, shineImpactHumanThalamus2023} to bias cortical computation towards regimes that are functionally suited to a given environmental situation.

The results of this work were obtained from simulations of a novel mathematical model of macroscale activity which incorporates both geometric and discrete connectomic structural mechanisms within a single modeling framework (provided in open code that accompanies this work).
The motivation to develop a new model is to improve upon existing models in literature, which face limitations in simultaneously capturing traveling wave dynamics along a continuous sheet and rapid communication dynamics through FNPs.
For example, one commonly used model is the neural mass model by Spiegler and colleagues, which simulates whole-brain stimulus-evoked responses  \cite{spieglerSilicoExplorationMouse2020, spieglerSelectiveActivationRestingState2016}.
Our model improves upon this model by generating finite-speed traveling waves (rather than infinite-speed diffusion), and enabling simulation of dynamics in multiple contexts other than stimulus-evoked responses.
Another commonly used model is the neural field model by Jirsa et al. \cite{qubbajNeuralFieldDynamics2007, jirsaNeuralFieldDynamics2009, jirsaSpatiotemporalPatternFormation2000} which simulates cortical activity on a one-dimensional continuum (line) with a single embedded FNP between two points.
While this model bears similarities to ours in generating traveling waves on a continuum, our model extends the one-dimensional geometry to a two-dimensional surface, to more accurately representing the geometry of the cortical sheet and enabling the model's structural connectivity to have multiple FNPs---consistent with the real cortex---without facing dynamical instabilities.
There is much scope for future work to further develop and refine models of brain activity that can simultaneously capture multiple mechanisms of cortico-cortical interactions, in order to better understand the types of interaction that shape cortical dynamics most strongly.

A notable limitation of this work is that our modeling results, and their interpretation, rest on specific assumptions underlying the model's construction.
One particular notable property of the model is its preservation of the total system activity over time (e.g., in `transporting' activity from one location to another, FNPs preserve total activity).
This property stands in contrast to other models, in which FNPs can propagate activity to multiple target locations and increase total activity, and nonlinearities in the local dynamics prevent dynamical instabilities arising from this increase \cite{decoRareLongrangeCortical2021}.
The validity of the activity-preserving assumption that underlies the construction of our model could be interrogated (or modified) in future work.
Another limitation of this work is that the model's behavior is validated through a qualitative comparison with observations with stimulus-response experiments.
Choices for model specification were made for the sake of simplicity and ease of demonstration and analysis, at the expense of empirical realism, including: setting the model's discrete connectome propagation parameters $\{c_m\}_m$ and $\{\tau_m\}_m$ to fixed values, and setting the cortical surface $\Omega$ to have periodic boundary conditions (and thus a toroidal geometry).
Also, when multiple FNPs were incorporated, the positions of each FNP were sampled under constraints of particular topological constraints observed in empirical literature, but not directly constrained by empirical data itself.
More quantitatively accurate model behavior can be achieved by setting $\Omega$ to a convoluted sphere per the real cerebral hemisphere \cite{gabayCorticalGeometryDeterminant2017}, aligning the wiring configurations of all FNPs with connectomic data as pursued in previous work \cite{decoRareLongrangeCortical2021, vohryzekHumanBrainDynamics2025}, and calibrating each $c_m$ and $\tau_m$ to neuroimaging data from different experimental paradigms.

In summary, we have developed a new model of cortical activity that allowed us to investigate the role of rapid, non-local projections in perturbing the otherwise geometrically constrained dynamics.
We find that the role of FNPs in shaping the dynamics depends strongly on the timescales resolved and the spatial precision of the driving input, suggesting that FNPs play a crucial role in shaping rapid stimulus-response dynamics, but can be approximately neglected in favor of simpler geometric rules on longer timescales or when driven by spatially diffuse drives.
These results provide a potential explanation for why some prior results demonstrate a strong role for FNPs in mediating non-local activations, while neglecting them in favor of an isotropic geometric rule becomes a stronger assumption in others.
Our paper is accompanied by open code to reproduce all results presented here \cite{GeometricFNPmodel}, and we hope it will be developed further in future work to better understand different mechanisms of cortical interaction.

\section{Acknowledgments}
We would like to thank Richa Phogat for helpful discussions regarding the activity preservation property of our model, and the assumption of zero time delays for our investigations, and Myuree Sivanesan for assistance with code reproducibility.
R.M. would like to thank the Australian Government Research Training Program (RTP) Scholarship for financial support.
B.D.F. acknowledges support from the Australian Research Council (FT240100418).

\newpage

\section{Supporting Information}

\renewcommand{\thefigure}{S\arabic{figure}}
\setcounter{figure}{0} 

\begin{figure}[!ht]
    \centering
    \includegraphics[width = \textwidth]{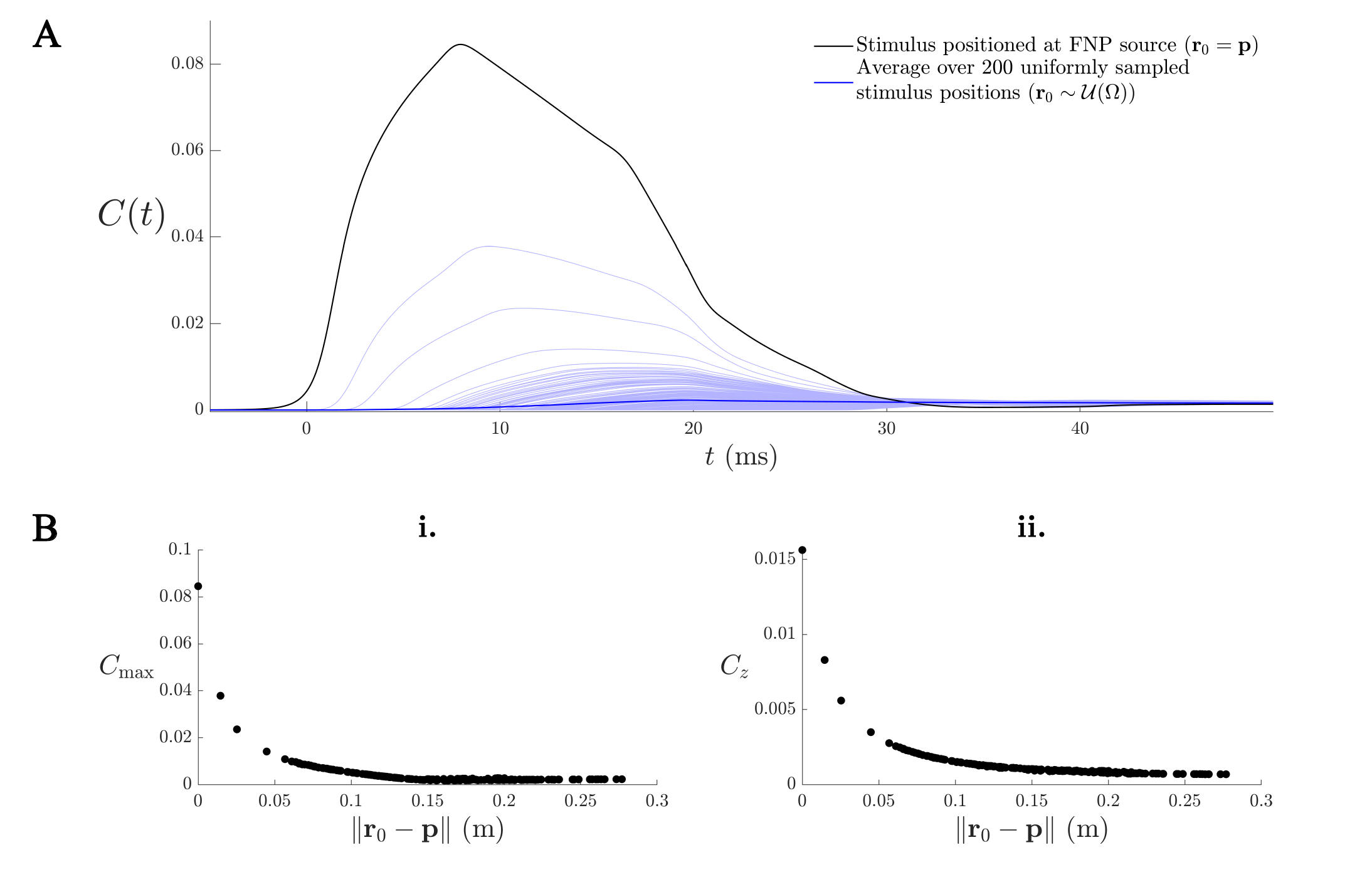}\qquad
    \captionsetup{font=small}
    \caption{
    \textbf{The short timescale influence of an FNP on stimulus-evoked dynamics is maximized when the stimulus is positioned at the source of the FNP, and drops off rapidly with increasing distance.}
    \textbf{A.} A plot of $C(t)$, the perturbation that an FNP from $\mathbf{p} = (3L/8,3L/8)$ to $\mathbf{q}=(5L/8,5L/8)$ induces on stimulus-evoked dynamics, for an ensemble of $200$ impulse stimulus positions $\mathbf{r}_0$ sampled uniformly on $\Omega$.
    The black curve is $C(t)$ when $\mathbf{r}_0 = \mathbf{p}$; i.e., when the stimulus is applied at the source of the FNP, which is identical to Fig.~\ref{fig:timescales}B. 
    \textbf{B.} For the same ensemble of stimulus positions, \textbf{i.} $C_{\max}$, the maximum attained $C(t)$ over time, is plotted against $\|\mathbf{r}_0 - \mathbf{p}\|$, the distance between the stimulus and the source of the FNP, and \textbf{ii.} $C_z$, which quantifies the FNP-induced perturbation on the evoked BOLD response, is plotted against $\|\mathbf{r}_0 - \mathbf{p}\|$.
    }
    \label{SI:fig:stim_position}
\end{figure}

\begin{figure}[!ht]
    \centering
    \includegraphics[width = \textwidth]{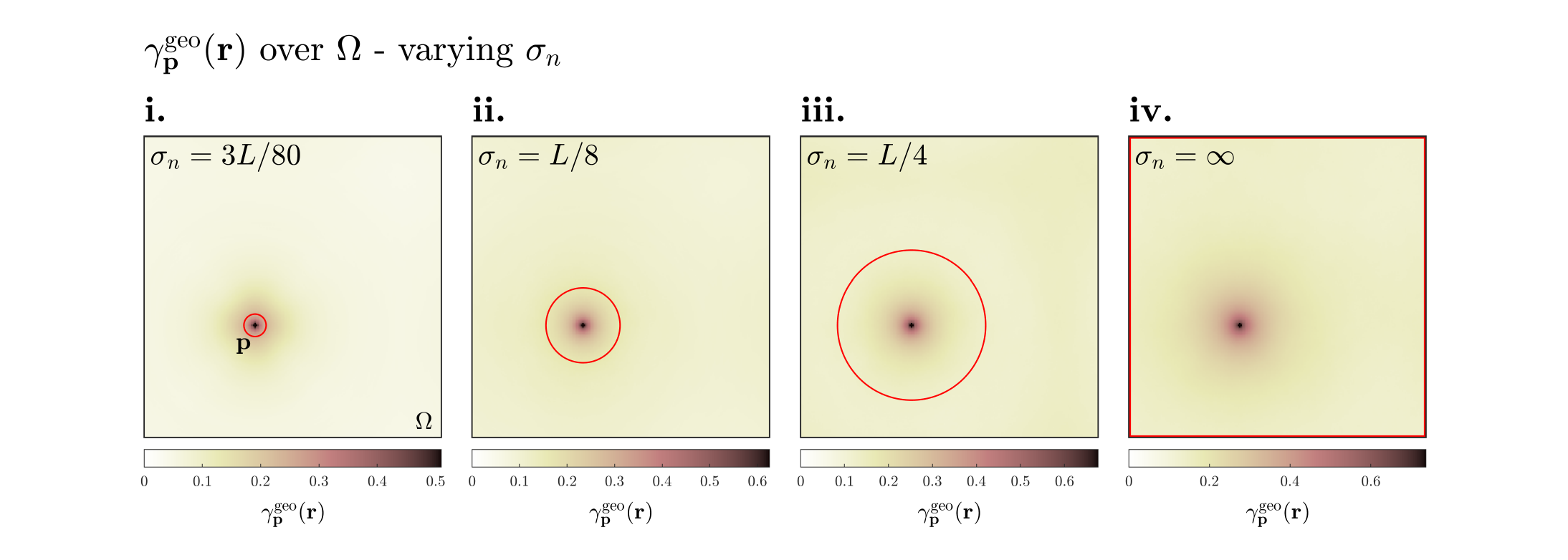}\qquad
    \captionsetup{font=small}
    \caption{
    \textbf{Spatial correlation function of the geometric model's noise-driven dynamics for different spatial profiles of noisy input.}
    A heatmap of $\gamma_\mathbf{p}^\text{geo}(\mathbf{r})$---the spatial correlation function of the geometric model with reference point $\mathbf{p}$---over $\Omega$ for four different spatial widths:
    \textbf{i.} $\sigma_n= 3L/80$,
    \textbf{ii.} $\sigma_n = L/8$,
    \textbf{iii.} $\sigma_n = L/4$, and
    \textbf{iv.} $\sigma_n = \infty$.
    The $\sigma_n = \infty$ case corresponds to a completely spatially unstructured input.
    }
    \label{SI:fig:spontaneous_homogeneous}
\end{figure}

\begin{figure}[!ht]
    \centering
    \includegraphics[width = \textwidth]{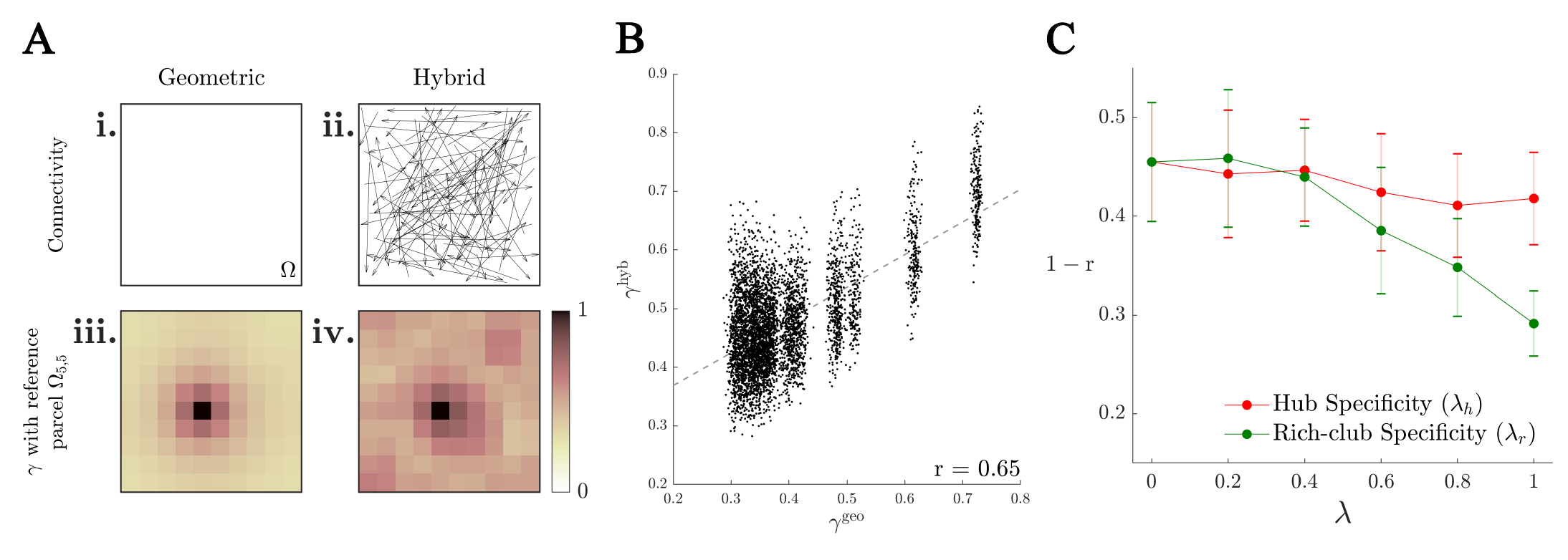}\qquad
    \captionsetup{font=small}
    \caption{
    \textbf{The influence of a connectome of $N=100$ FNPs on spontaneous dynamics is reduced by the incorporation of maximal hub specificity or maximal rich-club specificity.}
    \textbf{A.} For a given model, we computed its correlation structure as the set of Pearson correlations between activities of every pair of parcellated regions on $\Omega$.
    For our experiments, we divided $\Omega$ into $10 \times 10$ parcels: $\Omega_{ij} = (i-1, i]L/10 \times (j-1, j]L/10$ for $i,j = 1, ..., 10$.
    The activities of each parcel is then calculated as the pointwise activities $\phi(\mathbf{r}, t)$ integrated in space over the parcel.
    For both \textbf{i.} the geometric model, and \textbf{ii.} a hybrid model with a connectome of $N=100$ randomly sampled FNPs, a heatmap of the correlations between the example parcel $\Omega_{5,5}$ and all other $99$ parcels is shown in \textbf{iii.} and \textbf{iv.} respectively. 
    The correlation structures for both models are computed using the same algorithm to compute the spatial correlation function in Sec.~\nameref{SI:ResultsDetails}.
    \textbf{B.} We quantified the similarity between the spontaneous dynamics of the geometric and hybrid model as $\mathrm{r}$, the Pearson correlation between the $\textsuperscript{100}C_2$ elements of their parcellated correlation structures. 
    Here, the correlation structures of the geometric and hybrid model of \textbf{Aii} are compared.
    \textbf{C.} $1-\mathrm{r}$ quantifies the perturbation that the connectome of the hybrid model induces on the spontaneous dynamics.
    We compared the distribution of $1-\mathrm{r}$ over an ensemble of $20$ connectomes containing $N=100$ FNPs, for increasing hub specificity ($\lambda_h$) and rich-club specificity ($\lambda_r$).
    Each distribution is represented with errors bars of one standard deviation from the mean.
    The tolerance for the sampled connectomes was increased to $10^{-5}$ to speed up computation.
    }
    \label{SI:fig:connectome_nonrandom_corr}
\end{figure}

\paragraph*{S1}
\label{SI:modelderivation}
{\bf Detailed derivations of model.} 
In this appendix, we derive the equation of the mathematical model introduced and used in this work, describing the relevant variables and physiological mechanisms throughout.
The goal of this model is to approximate the spatiotemporal evolution of macroscale cortical activity---i.e., mean-field population-level activity on the cortical surface of a given cerebral hemisphere at millimeter lengthscales and above---from geometric propagation as traveling waves, concurrent with discrete connectomic propagation via fast-conducting non-local projections (FNPs) as rapid non-local interactions between specific populations on the surface.
For simplicity, we derive the equation for the case of a single FNP, as this idealized case can be easily extended to multiple FNPs  

The cortical surface (of either hemisphere) is mathematically represented as a closed continuous two-dimensional sheet denoted $\Omega$.
At each point $\mathbf{r}$ on $\Omega$, we assign local geometric and non-local connectomic propagation to two separate populations.
The locally propagating population, indexed $a=e$, has a spatially homogeneous and isotropic connectivity, and facilitates the geometric propagation.
In other words, this population connects every pair of populations on $\Omega$ with connectivity that is completely dependent on the distance between the two points. 
The non-local population, indexed $a=w$, is an FNP that connects population $\mathbf{a}$ to population $\mathbf{b}$, and facilitates rapid propagation between this pair only.
Finally, the inhibitory population, indexed $a=i$, only connects with other neurons in the same position. 
At a given point $\mathbf{r}$ over time $t$, we track two mean population-level activities for each population-type $a$: the mean firing rate of generated pulses (spikes) outgoing from the somas positioned at $\mathbf{r}$, $Q_a(\mathbf{r}, t)$, and the mean rate of pulses terminating at the synapses positioned at $\mathbf{r}$, $\phi_a(\mathbf{r}, t)$.
We assume that $\phi_a(\mathbf{r}, 0)=Q_a(\mathbf{r}, 0) = 0$ for all $a$, since the temporal dynamics of all variables are driven by external input rather than initial conditions.
The output variable of the model is $\phi_e(\mathbf{r}, t)$, per the construction of neural field models \cite{nunezNeocorticalDynamicsHuman1995}.

\begin{figure}[!ht]
    \centering
    \includegraphics[width = \textwidth]{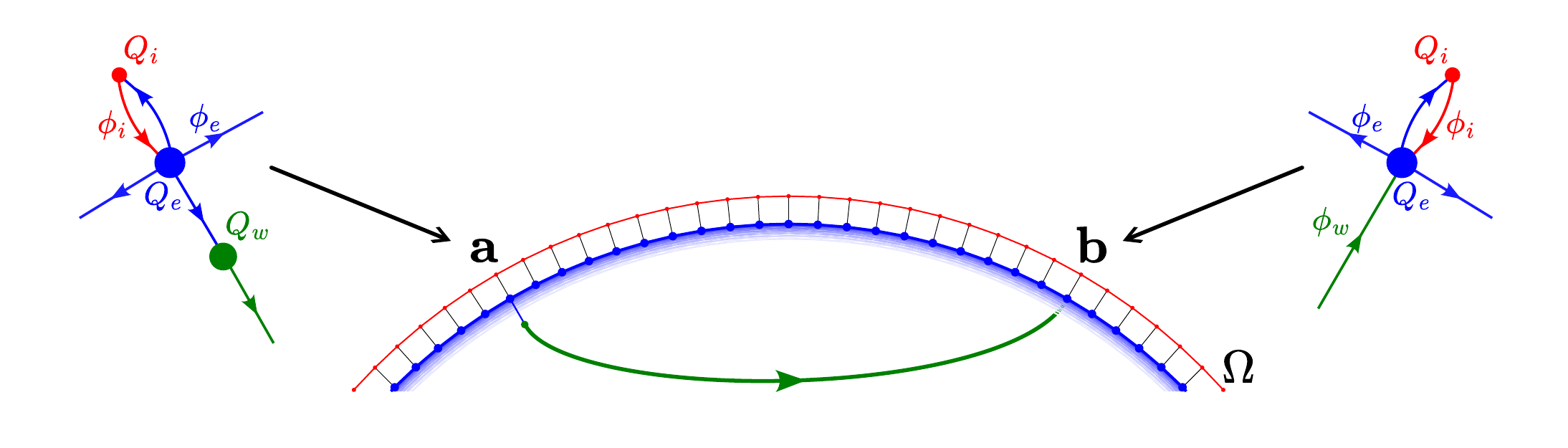}
    \captionsetup{font=small}
    \caption{
    \textbf{Schematic of physiological model.}
    }
    \label{SI:fig:physiologicalschematic}
\end{figure}

The propagator mechanisms of each population, which determine the influence of $Q_a$ on $\phi_a$ over space and time, is illustrated in Fig.~\ref{SI:fig:physiologicalschematic}.
Mathematically, the propagation of activity between two positions through the axons of population $a$ is represented by the Volterra integral equation
\begin{align}
\phi_a(\mathbf{r}, t) = \int_\Omega d^2 \mathbf{r}'\int_0^t \, dt' \, \, G_a(\mathbf{r}, t; \mathbf{r}', t') \,Q_a(\mathbf{r}', t') \text{ for all } (\mathbf{r}, t) \in \Omega \times [0, \infty)\,, \ a=e,i,w\,, \label{SI:eq:propagatorgeneral}
\end{align}
where the propagator of type $a$, $G_a$, is the proportion of a pulse originating at a soma of type $a$ at position $\mathbf{r}'$ and time $t'$, that terminates at a synapse of type $a$ at position $\mathbf{r}$ and time $t$.
Since all pulses terminate at a synapse, $G_a$ has unit integral over space and time.
Population $e$ propagates activity between any two points on $\Omega$, in an accordance to a distance-dependent connectivity rule.
Given that $\Omega$ is two-dimensional, we use the connectivity rule used in the two-dimensional neural field developed by Robinson et al. \cite{robinsonPropagationStabilityWaves1997}:
\begin{align}
    G_e(\mathbf{r}, t; \mathbf{r}', t') = \frac{\gamma e^{-\gamma (t - t')}}{2 \pi r} \frac{1}{\sqrt{r^2\gamma^2 (t - t')^2 - |\mathbf{r} - \mathbf{r}'|^2}} \Theta(r \gamma (t - t') - |\mathbf{r} - \mathbf{r}'|)\,, \label{SI:eq:propagator_e}
\end{align}
where $r$ is the spatial lengthscale of the connectivity rule, and $\gamma$ is the axonal propagation speed divided by $r$.
In contrast, population $w$ propagates activity strictly from point $\mathbf{a}$ to $\mathbf{b}$:
\begin{align}
G_w(\mathbf{r}, t; \mathbf{r}', t') = \delta(\mathbf{r} - \mathbf{b}) \, \delta(\mathbf{r}' - \mathbf{a}) \, \delta(t - t' - \tau), \label{SI:eq:propagator_w}
\end{align}
where $\tau$ is the conduction time for propagation in population $w$.
Finally, population $i$ propagates activity only to neurons in the same position:
\begin{align}
G_i(\mathbf{r}, t; \mathbf{r}', t') = \delta(\mathbf{r} - \mathbf{r}') \, \delta(t - t'). \label{SI:eq:propagator_i}
\end{align}
Substituting the propagators of each type in Eqs~\eqref{SI:eq:propagator_e},~\eqref{SI:eq:propagator_w},~\eqref{SI:eq:propagator_i} to Eq.~\eqref{SI:eq:propagatorgeneral} yields the equivalent differential equations:
\begin{align}
    &\left(1 + \frac{2}{\gamma}\frac{\partial}{\partial t} + \frac{1}{\gamma^2} \frac{\partial^2}{\partial t^2} - r^2 \nabla^2 \right) \phi_{e}(\mathbf{r}, t) = Q_e (\mathbf{r}, t) \,; \label{SI:eq:Qtophi_e} \\
    &\phi_w(\mathbf{r}, t) = \delta(\mathbf{r} - \mathbf{b}) \, Q_w(\mathbf{a}, t - \tau) \,; \label{SI:eq:Qtophi_w} \\
    &\phi_i(\mathbf{r}, t) = Q_i(\mathbf{r}, t) \,. \label{SI:eq:Qtophi_i}  
\end{align}

Complementing the propagator mechanisms of the model, are the synaptic mechanisms which map $\phi$ back to $Q$ at each position $\mathbf{r}$ and time $t$, illustrated in Fig.~\ref{SI:fig:physiologicalschematic}.
In summary, the synaptic activities of population type $e$ contribute to somatic activities of all population types ($\phi_e \to Q_e, Q_w, Q_i$), type $i$ contributes to types $e,i$ only ($\phi_i \to Q_e, Q_i$), and type $w$ contributes to type $e$ only ($\phi_w \to Q_e$).
In addition, external input $f$ contributes to the somatic activities of type $e$.
The choices of these interactions is to allow the total system to be reduced to a single equation, while capturing the following synaptic contributions: $\phi_e \to Q_w$ and $\phi_w \to Q_e$ for non-local connectomic propagation, $\phi_e \to Q_e$ for local geometric propagation, and $f \to Q_e$ for external input to drive the system. 
Mathematically, we used linear equations to represent the above synaptic mechanisms
\begin{align}
    &Q_e(\mathbf{r}, t) = G_{ee}(\mathbf{r}) \phi_e(\mathbf{r}, t) + G_{ei}(\mathbf{r}) \phi_i(\mathbf{r}, t) + G_{ew}(\mathbf{r}) \phi_w(\mathbf{r}, t) + f(\mathbf{r}, t) \,; \label{SI:eq:phitoQ_e} \\
    &Q_w(\mathbf{r}, t) = G_{we}(\mathbf{r}) \phi_e(\mathbf{r}, t) \,; \label{SI:eq:phitoQ_w} \\ 
    &Q_i(\mathbf{r}, t) = G_{ie}(\mathbf{r}) \phi_e(\mathbf{r}, t) + G_{ii}(\mathbf{r}) \phi_i(\mathbf{r}, t) \,. \label{SI:eq:phitoQ_i} 
\end{align}
In the equations above, each gain term $G_{ab}(\mathbf{r})$ is the change in $Q_a$ at position $\mathbf{r}$, per unit increase in $\phi_b$ at the same position and time.
$G_{ab}$ is positive for $b = e, w$ and negative for $b = i$.
Physiologically, this relationship arises from the pulses of the synapses of $b$ being `transferred' to the dendrites of $a$ through neurotransmitter mechanisms, before arriving at the somas of $a$  \cite{robinsonPropagatorTheoryBrain2005}.
The equations neglect the temporal spreading that arises from conducting delays within the dendritic trees of individual neurons, as these delays do not differ between locally and non-locally propagating populations ($e, w$), thus out of the model's scope.
The equations also assume the transfer of pulses is linear,  which is a reasonable approximation if the model's variables are defined as first-order deviations about an assumed steady state \cite{robinsonPropagationStabilityWaves1997}.
The above Eqs~\eqref{SI:eq:Qtophi_e},~\eqref{SI:eq:Qtophi_w},~\eqref{SI:eq:Qtophi_i},~\eqref{SI:eq:phitoQ_e},~\eqref{SI:eq:phitoQ_w},~\eqref{SI:eq:phitoQ_i}, when combined, reduces the system to a single non-local partial differential equation, with variable $\phi = \phi_e(\mathbf{r}, t)$, and zero initial conditions.
\begin{align}
    &\left(1 - \tilde{G}_e(\mathbf{r}) + \frac{2}{\gamma}\frac{\partial}{\partial t} + \frac{1}{\gamma^2} \frac{\partial^2}{\partial t^2} - r^2 \nabla^2 \right) \phi(\mathbf{r}, t) =  \tilde{G}_w(\mathbf{r}) \phi(\mathbf{a}, t - \tau) + f(\mathbf{r}, t)  \text{, where}  \label{SI:eq:modelintermediate} \\
    &\qquad \tilde{G}_e(\mathbf{r}) = G_{ee}(\mathbf{r}) + \frac{G_{ei}(\mathbf{r}) G_{ie}(\mathbf{r})}{1 - G_{ii}(\mathbf{r})}, \text{ and } \tilde{G}_w(\mathbf{r}) = G_{ew}(\mathbf{b}) G_{we}(\mathbf{b}) \delta(\mathbf{r}-\mathbf{b}) \,; \nonumber \\
    &\phi(\mathbf{r}, 0) = \dot \phi(\mathbf{r}, 0) = 0 \,. \label{SI:eq:initconditions}
\end{align}
Finally we aim to calibrate the effective gain terms $\tilde{G}_e(\mathbf{r}), \tilde{G}_w(\mathbf{r})$ in space, to satisfy the property that, for any given input $f(\mathbf{r}, t)$, the total activity in the system integrated over space and time is preserved after the addition of population $w$.
We let $\phi_0(\mathbf{r}, t)$ be the solution to the system in the absence of population $w$.
Then, we have the constraint
\begin{align}
    \int_0^\infty dt \int_\Omega d^2 \mathbf{r} \ \phi(\mathbf{r}, t) = \int_0^\infty dt \int_\Omega d^2 \mathbf{r} \ \phi_0(\mathbf{r}, t). \label{SI:eq:energypreservation}
\end{align}

In the case of the geometric model, where population $w$ is absent, $\tilde{G}_e(\mathbf{r})$ is spatially uniform, and $\tilde{G}_w(\mathbf{r})=0$ for all $\mathbf{r}$.
We let $\nu_0$ be the constant value of $\tilde{G}_e(\mathbf{r})$ in the absence of population $w$.
Thus $\phi_0$ is the solution to the differential equation
\begin{align}
    & \left(1 - \nu_0 + \frac{2}{\gamma}\frac{\partial}{\partial t} + \frac{1}{\gamma^2} \frac{\partial^2}{\partial t^2} - r^2 \nabla^2 \right) \phi_0(\mathbf{r}, t) = f(\mathbf{r}, t)\,; \label{SI:eq:geometricmodelintermediate} \\
    & \phi_0(\mathbf{r}, 0) = \dot \phi_0(\mathbf{r}, 0) = 0 \,. \label{SI:eq:geometricinitconditions}
\end{align}
The local synaptic mechanisms of population $e$ in our model is identical at all points on $\Omega$ except $\mathbf{a}$, where synaptic activities of type $e$ contribute to population type $w$.
Accordingly, we parameterize the effective gain term $\tilde{G}_e(\mathbf{r})$ to differ from $\nu_0$ at point $\mathbf{a}$ only, i.e., 
\begin{align}
    \tilde{G}_e(\mathbf{r}) = \nu_0 - c\,\delta(\mathbf{r} - \mathbf{a}),  \label{SI:eq:Ge}
\end{align} 
for some constant $c > 0$.
Under the initial conditions in Eqs~\eqref{SI:eq:initconditions},~\eqref{SI:eq:geometricinitconditions}, and the fact that $\int_\Omega d^2 \mathbf{r} \ \nabla^2 \phi(\mathbf{r}, t) = \int_\Omega d^2 \mathbf{r} \ \nabla^2 \phi_0(\mathbf{r}, t) = 0$ for all $t$ since $\Omega$ is a closed surface, integrating Eqs~\eqref{SI:eq:modelintermediate},~\eqref{SI:eq:geometricmodelintermediate} over space and time, and then applying constraint Eq.~\eqref{SI:eq:energypreservation}, yields equality $G_{ew}(\mathbf{b}) G_{we}(\mathbf{b})=c$, implying
\begin{align}
    \tilde{G}_w(\mathbf{r}) = c \,\delta(\mathbf{r} - \mathbf{b}). \label{SI:eq:Gw}
\end{align}
Substituting Eqs~\eqref{SI:eq:Ge} and \eqref{SI:eq:Gw} into Eq.~\eqref{SI:eq:modelintermediate} yields the simplified equation:
\begin{align*}
    &\left( 1 - \nu_0 + \frac{2}{\gamma}\frac{\partial}{\partial t} + \frac{1}{\gamma^2} \frac{\partial^2}{\partial t^2} - r^2 \nabla^2 \right) \phi(\mathbf{r}, t) \\
    &\qquad = -c \, \delta(\mathbf{r}-\mathbf{a}) \, \phi(\mathbf{a}, t) + c \, \delta(\mathbf{r} - \mathbf{b}) \, \phi(\mathbf{a}, t - \tau) + f(\mathbf{r}, t) \,, \\
    &\phi(\mathbf{r}, 0) = \dot \phi(\mathbf{r}, 0) = 0 \,.
\end{align*}
Finally, we generalize the equation to multiple FNPs, where the $m^\text{th}$ FNP has source and term $\mathbf{a}_m, \mathbf{b}_m$, and connectivity parameter $c_m$. 
Accordingly, we attain the equation:
\begin{align}
    &\mathcal{D}_t \phi(\mathbf{r}, t) - r^2 \nabla^2(\mathbf{r}, t) \phi - (C \circ \phi)(\mathbf{r}, t) = f(\mathbf{r}, t) \,, \text{ where }\label{SI:eq:modeldetailed} \\
    &\qquad \mathcal{D}_t \phi(\mathbf{r}, t) = \left(1 - \nu_0 + \frac{2}{\gamma}\frac{\partial}{\partial t} + \frac{1}{\gamma^2} \frac{\partial^2}{\partial t^2} \right) \phi(\mathbf{r}, t) \,, \label{SI:eq:differentialoperator} \\
    &\qquad \left(C \circ \phi\right)(\mathbf{r}, t) = \sum_m c_m \big[ \delta(\mathbf{r} - \mathbf{b}_m) \phi(\mathbf{a}_m, t - \tau_m) - \delta(\mathbf{r} - \mathbf{a}_m) \phi(\mathbf{a}_m, t) \big] \,; \label{SI:eq:nonlocalterm} \\
    &\phi(\mathbf{r}, 0) = \dot \phi(\mathbf{r}, 0) = 0 \,. \nonumber
\end{align}

\paragraph*{S2}
\label{SI:NumericalTreatment}
{\bf Details of Numerical Treatment of Model.}
In this appendix, we detail the numerical methods used to solve the differential equation of the model used in this work.
The equation can be found in Eq.~\eqref{SI:eq:modeldetailed},~\eqref{SI:eq:differentialoperator},~\eqref{SI:eq:nonlocalterm}.
The methods used below also assume the particular model specifications used in this work: that the spatial domain is a square of length $L$ with periodic boundary conditions ($\Omega = (0, L]^2$), and the conduction time of the $m^\text{th}$ FNP is zero ($\tau_m=0$) for all $m$.

For accessibility, we first rewrite the differential equation to separate the temporal derivative terms from other terms:
\begin{align}
    &\phi(\mathbf{r}, t) + \frac{2}{\gamma} \dot \phi(\mathbf{r}, t) + \frac{1}{\gamma^2} \ddot \phi(\mathbf{r}, t) = P(\mathbf{r}, t) \,,  \text{ for all } (\mathbf{r}, t) \in \Omega \times [0, \infty)\,, \text{ where } \label{SI:eq:modelrearranged} \\
    &\qquad P(\mathbf{r}, t) = \nu_0 \phi(\mathbf{r}, t) + r^2 \nabla^2 \phi(\mathbf{r}, t) + (C \circ \phi)(\mathbf{r}, t) + f(\mathbf{r}, t) \,; \nonumber \\
    &\phi(\mathbf{r}, 0) = \dot \phi(\mathbf{r}, 0) = 0 \,. \nonumber
\end{align}
From $\Omega$, we uniformly partitioned a $N_x \times N_x$ computational grid of points with grid spacing $\Delta x = L / N$: 
\begin{align}
    \Omega^* = \{\mathbf{r}_{ij}:i, j=1, \ldots, N_x\} \subset \Omega, \text{ where } \mathbf{r}_{ij}=(i \Delta x, j \Delta x)\,.\label{SI:eq:computationalgrid}
\end{align}
For the sake of computer simulation, the time domain is truncated to a bounded interval $[0, T]$, from which we uniformly partitioned $1+N_t$ time points $\{t_n = n\Delta t: n = 0, 1, \ldots N_t \}$ with a time step $\Delta t = T/N_t$.

We define $\phi_{ij}^n = \phi(\mathbf{r}_{ij}, t_n)$ and $P_{ij}^n = P(\mathbf{r}_{ij}, t_n)$ as the numerical solutions at each grid point and time step.
At each time step $n$, the numerical scheme we outline below computes $P_{ij}^n$ for $i, j$, and uses this value to compute $\phi_{ij}^{n+1}$ for all $i, j$.

$P_{ij}^n$ is computed on the computational grid as the sum of: $r^2 \nabla^2(\mathbf{r}_{ij}, t_n)$, the local term; $(C \circ \phi)(\mathbf{r}_{ij}, t_n)$, the non-local term, and; $f(\mathbf{r}_{ij}, t_n)$, the external input.
The first term, the local term, is approximated by the centered finite difference (five-point stencil), while incorporating the periodic boundary conditions of $\Omega$:
\begin{align}
   &r^2 \nabla^2 \phi(\mathbf{r}_{ij}, t_n) \approx B_{ij}^n = \frac{r^2}{(\Delta x)^2} \left(\phi_{i + 1, j}^n + \phi_{i, j + 1}^n + \phi_{i - 1, j}^n + \phi_{i, j - 1}^n - 4\phi_{ij}^n\right)\,, \label{SI:eq:laplaciannumericaltreatment} \\
   &\qquad \phi_{N+1,j}^n = \phi_{1,j}^n, \ \phi_{0,j}^n = \phi_{N,j}^n, \ \phi_{i,N+1}^n = \phi_{i,1}^n, \ \phi_{i,0}^n = \phi_{i,N}^n\,. \nonumber
\end{align}

The second term, the non-local term, is computed as
\begin{align*}
    \left(C \circ \phi\right)(\mathbf{r}_{ij}, t_n) = \sum_m c_m \big[ \delta(\mathbf{r}_{ij} - \mathbf{b}_m) - \delta(\mathbf{r}_{ij} - \mathbf{a}_m)\big] \phi(\mathbf{a}_m, t_n).
\end{align*}
since $\tau_m=0$ for all $m$. 
Due to discontinuities of the delta terms $\delta(\mathbf{r}_{ij}-\mathbf{a}_m), \delta(\mathbf{r}_{ij}-\mathbf{b}_m)$ at $\mathbf{a}_m, \mathbf{b}_m$ respectively, and the possibility that the pointwise evaluation $\phi(\mathbf{a}_m, t_n)$ may not be attainable due to discretization, $(C \circ \phi)(\mathbf{r}_{ij}, t_n)$ is numerically computed with mollifiers, which are smooth functions that approximate the delta function. 
We first rewrite the term as
\begin{align*}
\left(C \circ \phi\right)(\mathbf{r}_{ij}, t_n) = \lim_{\epsilon \to 0} \sum_m c_m \big[ k_\epsilon(\mathbf{r}_{ij} - \mathbf{b}_m) - k_\epsilon(\mathbf{r}_{ij} - \mathbf{a}_m) \big] \int_\Omega d^2 \mathbf{r}' k_\epsilon(\mathbf{r}' - \mathbf{a}_m) \phi(\mathbf{r}', t_n)
\end{align*}
where $k_\epsilon$ is a mollifier of parameter $\epsilon$ with unit integral over $\Omega$. 
For our experiments we approximate $(C \circ \phi)(\mathbf{r}_{ij}, t_n)$ by setting the mollifers $k_\epsilon$ as Gaussian in space, and approximating the resultant integral by the Riemann sum expression:
\begin{align}
    &\left(C \circ \phi\right)(\mathbf{r}_{ij}, t_n) \approx C_{ij}^n =   \sum_{m=1} \frac{c_m}{(\Delta x)^2} \Big[ w_{ij}^{\mathbf{b}_m} - w_{ij}^{\mathbf{a}_m} \Big] \sum_{i', j'} w_{i'j'}^{\mathbf{a}_m} \phi_{i'j'}^n \text{, where }\label{SI:eq:nonlocalnumericaltreatment}\\ 
    &\qquad w_{ij}^{\mathbf{a}_m} \propto \exp\left(\frac{- \lVert\mathbf{r}_{ij} -\mathbf{a}_m\rVert^2}{2 \epsilon^2} \right) , \ w_{ij}^{\mathbf{b}_m} \propto \exp\left(\frac{- \lVert\mathbf{r}_{ij} -\mathbf{b}_m\rVert^2}{2 \epsilon^2} \right), \nonumber \\ &\qquad \sum_{i,j} w_{ij}^{\mathbf{a}_m} = \sum_{i,j} w_{ij}^{\mathbf{b}_m} = 1\,. \nonumber
\end{align}

The third term, the external input term, is computed differently depending on whether the spatiotemporal profile of $f(\mathbf{r}, t)$ is an impulse-like input ($f^\text{stim}(\mathbf{r}, t)$ in Eq.~\eqref{eq:impulsestimulus}) used to simulate stimulus-evoked responses, or a continuous space-time stochastic process ($f^\text{noise}(\mathbf{r}, t)$ in Eq.~\eqref{eq:whitenoise}) used to simulate noise-driven activity.
In the first case, the input term is pointwise evaluated at each grid point and timestep:
\begin{align}
    &f(\mathbf{r}_{ij}, t_n) \approx f_{ij}^{n, \text{stim}} \propto \exp\left(\frac{-\lVert\mathbf{r}_{ij} -\mathbf{r}_{os}\rVert^2}{2 \sigma_x^2} \right) \exp\left(\frac{-(t_n -t_{os})^2}{2 \sigma_t^2} \right) \,, \label{eq:inputimpulsenumericaltreatment} \\ 
    &\qquad \sum_n \sum_{i,j} f_{ij}^{n,\text{stim}} = \frac{1}{(\Delta t) (\Delta x)^2} \,. \nonumber
\end{align}
In second case, the white noise process $\dot W$ is treated as a piecewise constant random process \cite{langtangen2016finite}, whereas the deterministic function $g(\mathbf{r})$ is pointwise evaluated at each grid point:
\begin{align}
    &f(\mathbf{r}_{ij}, t_n) \approx f_{ij}^{n, \text{noise}} = g_{ij} \frac{1}{\sqrt{\Delta t} \Delta x} \eta_{ij} \,, \label{eq:inputnoisenumericaltreatment} \\ 
    &\qquad g_{ij} = g(\mathbf{r}_{ij}) \,, \ \eta_{ij} \sim \mathcal{N}(0, 1)\,, \nonumber
\end{align}
where $\mathcal{N}$ is the normal distribution.

The above Eqs~\eqref{SI:eq:laplaciannumericaltreatment},~\eqref{SI:eq:nonlocalnumericaltreatment},~\eqref{eq:inputimpulsenumericaltreatment},~\eqref{eq:inputnoisenumericaltreatment} are used to compute $P_{ij}^n$. 
This value is then used to compute $\phi_{ij}^{n+1}$ from $\phi_{ij}^n$ and $\phi_{ij}^{n-1}$.
In this computation, the temporal derivatives in Eq.~\eqref{SI:eq:modelrearranged} are approximated by centered finite differences:
\begin{align*}
    \ddot \phi(\mathbf{r}_{ij}, t_n) \approx \frac{\phi_{ij}^{n+1} - 2\phi_{ij}^n + \phi_{ij}^{n-1}}{(\Delta t)^2} \,, \ 
    \dot \phi(\mathbf{r}_{ij}, t_n) \approx \frac{\phi_{ij}^{n+1} - \phi_{ij}^{n-1}}{2 (\Delta t)} \,.
\end{align*}
In the case where $n=0$, we substitute the initial conditions $\phi(\mathbf{r}, 0) = \dot \phi(\mathbf{r}, 0) = 0$ into the above finite difference approximations to obtain the approximations $\ddot \phi(\mathbf{r}_{ij}, t_0) = 2\phi_{ij}^1(\Delta t)^{-2}\,, \ \dot \phi(\mathbf{r}_{ij}, t_0) = 0$\,.

Put together, the numerical scheme used for the model is shown below: 
\begin{align}
    &\alpha = \frac{(\gamma \Delta t)^2}{2}; \ 
    \beta_1 = \frac{(\gamma \Delta t)^2}{\gamma \Delta t + 1}; \ 
    \beta_2 = \frac{2 - (\gamma \Delta t)^2}{\gamma \Delta t + 1}; \ 
    \beta_3 = \frac{\gamma \Delta t - 1}{\gamma \Delta t + 1}; \label{SI:eq:numericaltreatment} \\
    &\forall i,j = 1, \ldots N_x: \phi_{ij}^0 = 0; \nonumber \\
    &\forall n = 0, 1, \ldots, N_t-1: \nonumber \\
    &\qquad \forall i,j = 1, \ldots N_x: P_{ij}^n = \nu_0 \phi_{ij}^n + B_{ij}^n + C_{ij}^n + f_{ij}^n; \ ~\eqref{SI:eq:laplaciannumericaltreatment},~\eqref{SI:eq:nonlocalnumericaltreatment},~\eqref{eq:inputimpulsenumericaltreatment},~\eqref{eq:inputnoisenumericaltreatment} \nonumber \\
    &\qquad \forall i,j = 1, \ldots N_x: \phi_{ij}^{n+1} = \begin{cases}
        \alpha P_{ij}^0, \ n = 0, \\
        \beta_1 P_{ij}^n + \beta_2 \phi_{ij}^n + \beta_3 \phi_{ij}^{n-1}, \ n \geq 1\,.
    \end{cases} \nonumber
\end{align}

The values of the quantities introduced above for numerical treatment, $N_x, T, N_t, \epsilon$, used for our investigations are shown in Table~\ref{SI:tab:numericaltreatment_values}.
$N_x$ is fixed to the value $200$ to finely resolve radial traveling waves at millimeter order spatial resolution ($\Delta x = \SI{0.4}{\m}/200=\SI{2}{\mm}$).
The default value of $T$ was fixed to $\SI{70}{\ms}$ to simulate the time-course of stimulus-evoked responses over short millisecond timescales.
$N_t$ is fixed to the value $988$ to satisfy the Courant--Friedrichs--Lewy condition of the two-dimensional wave equation ($\Delta x / \Delta t > r\gamma \sqrt{2}$) \cite{linesRecipeStabilityFinitedifference1999}.
$\epsilon$ was fixed to $\SI{0.002}{\m}$, which was small enough for the mollifer $k_\epsilon$ to mimic the delta functional without encountering spurious oscillations in the spontaneous experiments in Fig.~\ref{fig:spontaneous}.

\begin{table}[!ht]
\centering
\caption{
{\bf Quantities for numerical treatment}
}
\begin{tabular}{|l|l|}
\hline
\bf{Quantity} & \bf{Value}\\ \thickhline
$N_x$ & $200$ \\ \hline
$T$ & $\SI{0.07}{\s}$ \\ \hline
$N_t$ & $988$ \\ \hline
$\epsilon$ & $\SI{0.002}{\m}$\\ \hline
\end{tabular}
\label{SI:tab:numericaltreatment_values}
\end{table}

\paragraph*{S3}
\label{SI:ResultsDetails}
{\bf Details of Computations in Results.}

\textbf{Computation of BOLD responses.}
Given the model's stimulus-evoked response $\phi(\mathbf{r}, t)$ to a impulse stimulus under a given structural connectivity, the BOLD response (BOLD signal observation of the stimulus-evoked response) is approximated by $z(\mathbf{r})$, the zero temporal frequency component of $\phi(\mathbf{r}, t)$:
\begin{equation}
    z(\mathbf{r}) = \mathcal{F}[\phi](\mathbf{r}, \omega = 0) = \int_0^\infty dt \ \phi(\mathbf{r}, t). \label{SI:eq:bold}
\end{equation}
Here, $\mathcal{F}$ is the Fourier transform as an operator on $\phi$ as a function of $t$ only.
$z(\mathbf{r})$ is computed numerically at each grid point $\mathbf{r}_{ij}$ by computing the matrix $Z = \left(z(\mathbf{r}_{ij})\right)_{ij}$. 
$Z$ is computed as the matrix series
\begin{equation}
    Z = \Delta t \,\sum_{n=1}^\infty \Phi[n] \,,
\end{equation}
where $\Phi[n] =\left(\phi_{ij}^n\right)_{ij}$.
To compute the above series, we iteratively compute the partial sum 
\begin{align*}
    Y[m] = \Delta t \,\sum_{n=0}^{m\cdot N_t}  \Phi[n] \text{ for } m = 1, 2, \ldots
\end{align*}
We stop the iteration when the cosine distance between the vectors  $\mathrm{vec}(Y[m]), \mathrm{vec}(Y[m-1])$, the vectorizations of the matrices $Y[m], Y[m-1]$, is less than the threshold $\mathrm{tol}=10^{-5}$. 
The last computed $Y[m]$ is taken as an approximate scalar multiple of $Z$, since the cosine distance between $\mathrm{vec}(Y[m])$ and $\mathrm{vec}(Z)$ is close to zero.
To compute $Z$ from $Y_m$, we use the fact that, by integrating both sides of \eqref{SI:eq:modeldetailed} over both space $\Omega$ and time $[0, \infty)$, one obtains
\begin{align*}
    (1-\nu_0) \left(\int_\Omega d^2\mathbf{r} \ z(\mathbf{r}) \right) &= \int_0^\infty dt \int_\Omega d^2 \mathbf{r} \ \delta(\mathbf{r} - \mathbf{r}_0) \delta(t - t_0) = 1 \\
    \implies \int_\Omega d^2\mathbf{r} \ z(\mathbf{r}) &= (1-\nu_0)^{-1}.
\end{align*}
Accordingly, $Z$ is computed by scalar mutliplying $Y[m]$ so that the sum of all elements in $Z$ is $(\Delta x)^{-2} (1 - \nu_0)^{-1}$.

\textbf{Computation of Cosine Dissimilarity Metrics.}
The cosine dissimilarity between the stimulus-evoked dynamics of the geometric model and any hybrid model containing additional FNPs, in response to the same impulse stimulus, is defined by
\begin{equation}
    C_\phi(t) = 1 - \frac{\int_\Omega d^2 \mathbf{r} \ \phi(\mathbf{r}, t) \phi_0(\mathbf{r}, t)}
    {\sqrt{\int_\Omega d^2 \mathbf{r} \left(\phi(\mathbf{r}, t)\right)^2} \sqrt{\int_\Omega d^2 \mathbf{r} \left(\phi_0(\mathbf{r}, t)\right)^2}}\,, \label{SI:eq:cosinedissimilarity}
\end{equation}
where $\phi_0$ and $\phi$ are the geometric model's and hybrid model's activity respectively.
Similarly, the cosine dissimilarity between the evoked BOLD response of the geometric model and any hybrid model containing additional FNPs, in response to the same impulse stimulus, is defined by
\begin{equation}
    C_z = 1 - \frac{\int_\Omega d^2 \mathbf{r} \ z(\mathbf{r}) z_0(\mathbf{r})}
    {\sqrt{\int_\Omega d^2 \mathbf{r} \left(z(\mathbf{r})\right)^2} \sqrt{\int_\Omega d^2 \mathbf{r} \left(z_0(\mathbf{r})\right)^2}}\,, \label{SI:eq:cosinedissimilaritybold}
\end{equation}
where $z_0$ and $z$ are the geometric model's and hybrid model's BOLD responses respectively.
$C_\phi(t)$ is used in our investigations to quantify the perturbation of an FNP, or a connectome of multiple FNPs on the model's stimulus-evoked dynamics over time, and $C_z$ quantifies the perturbation on the evoked BOLD response.
To compute $C_\phi(t_n)$, where $t_n=n\Delta t$ is a timestep in the simulation, we computed the cosine distance between the vectors $\mathrm{vec}(\Phi[n])$ and $\mathrm{vec}(\Phi_0[n])$, where $\Phi[n] = \left(\phi(\mathbf{r}_{ij}, t_n)\right)_{ij}, \Phi_0[n] = \left(\phi_0(\mathbf{r}_{ij}, t_n)\right)_{ij}$.
Similarly, $C_z$ is computed as the cosine distance between $\mathrm{vec}(Z)$ and $\mathrm{vec}(Z_0)$, where $Z = \left(z(\mathbf{r}_{ij})\right)_{ij}, Z_0 = \left(z_0(\mathbf{r}_{ij})\right)_{ij}$.

\textbf{Computation of Spatial Correlation Function.}
The spatial correlation function with reference point $\mathbf{p}$, denoted $\gamma_\mathbf{p}:\Omega\to[-1, 1]:\mathbf{r}\to \mathrm{corr}(\phi(\mathbf{p},t),\phi(\mathbf{r}, t))$, returns the Pearson correlation between the model's noise-driven activity at point $\mathbf{p}$ and any input point $\mathbf{r}$ on $\Omega$.
In our investigations, we numerically compute $\gamma_\mathbf{p}$ evaluated at all grid points $\mathbf{r}_{ij}$ by computing the matrix $\Gamma = \left(\gamma_\mathbf{p}(\mathbf{r}_{ij})\right)_{ij}$.
Each element of $\Gamma$ is computed as
\begin{align*}
    &\gamma_\mathbf{p}(\mathbf{r}_{ij}) = \lim_{m\to\infty} r_{ij}[m], \\
    &\qquad r_{ij}[m] = \frac{\frac{\sum_{n=1}^{m\cdot N_t}\phi_\mathbf{p}^n \phi_{ij}^n}{m \cdot N_t} - \frac{\sum_{n=1}^{m\cdot N_t}\phi_\mathbf{p}^n}{m \cdot N_t} \cdot \frac{\sum_{n=1}^{m\cdot N_t} \phi_{ij}^n}{m \cdot N_t}}{\sqrt{ \frac{\sum_{n=1}^{m\cdot N_t}\left(\phi_\mathbf{p}^n \right)^2}{m \cdot N_t} - \left( \frac{\sum_{n=1}^{m\cdot N_t}\phi_\mathbf{p}^n}{m \cdot N_t} \right)^2} \sqrt{ \frac{\sum_{n=1}^{m\cdot N_t}\left(\phi_{ij}^n \right)^2}{m \cdot N_t} - \left( \frac{\sum_{n=1}^{m\cdot N_t}\phi_{ij}^n}{m \cdot N_t} \right)^2 }},
\end{align*}
where $\phi_{\mathbf{p}}^n$ is the simulated time series $\phi(\mathbf{r}, t_n)$ at point $\mathbf{p} = (0.15,0.15)$.
For grid size $N_x = 200$ and domain length $L = 0.4$, $\phi_{\mathbf{p}}^n$ is the time series $\phi_{75,75}^n$. 
Iteratively with $m = 1,2,\ldots$, we compute the matrix $R[m] = (r_{ij}[m])_{ij}$.
We stop the iteration when the correlation between the vectors $\mathrm{vec}(R[m]), \mathrm{vec}(R[m-1])$ is less than the threshold $\mathrm{tol} = 10^{-7}$.
The last attained $R[m]$ is then taken as the numerical computation of $\Gamma$.

\textbf{Sampling of Connectomes with non-random properties}
Connectomes constrained by non-random properties (exponential distance rule, hub specificity and rich-club specificity) are sampled using a rejection-sampling algorithm.
The source and target point of each FNP is sampled uniformly and independently on $\Omega$, but is accepted only if the sampled source and target passes a acceptance-rejection test tailored to the non-random property under investigation; otherwise the source and target is sampled again.
Once the sampled source and target is accepted, the next FNP is sampled in a similar manner.
We emphasize, however, that each connectome we generate is constrained by one of the three non-random properties only.

\textbf{Acceptance-rejection Test for Exponential Distance Rule constraint. }
To sample an FNP with parameter $\lambda_e$, a source $\mathbf{r}_1$ and target $\mathbf{r}_2$ is first sampled uniformly from $\Omega$.
A value $u$ is then sampled from the $\mathrm{Uniform}[0,1]$ distribution.
$\mathbf{r}_1$ and $\mathbf{r}_2$ are accepted if $u<\exp(-100\lambda_e l)$, where $\lambda_e \in [0, 1]$ is the selected decay parameter, and $l$ is the Euclidean distance between the source and target (with periodic boundary conditions).
Otherwise, $\mathbf{r}_1$, $\mathbf{r}_2$ and $u$ are resampled.
In this way, the probability that a FNP exists with source $\mathbf{r}_1$ and target $\mathbf{r}_2$, for any two given points $\mathbf{r}_1$ and $\mathbf{r}_2$, is 
\begin{align}
\mathbb{P}\left(\text{FNP $\mathbf{r}_1 \to \mathbf{r}_2$} \ | \ \mathbf{r}_1, \mathbf{r}_2\right) = \exp \left( -100\lambda_e \| \mathbf{r}_1 - \mathbf{r}_2 \| \right). \label{SI:eq:edr}
\end{align}

\textbf{Acceptance-rejection Test for Hub Specificity constraint. }
To sample an FNP with parameter $\lambda_h$, a value $u$ is first sampled from the $\mathrm{Uniform}[0,1]$ distribution.
If $u < \lambda_h$, the source and target points of the FNP are repetitively sampled uniformly from $\Omega$ and accepted when the FNP is hub-connecting, i.e., the FNP connects a point inside any one hub with a point outside that hub.
Otherwise, if $u \geq \lambda_h$, the source and target points are sampled once and accepted. 
In this way, the probability that any given FNP is hub-connecting is 
\begin{align}
    \mathbb{P}\left(\text{FNP is hub-connecting}\right) = (1 - \lambda_h)p_h + \lambda_h, \label{SI:eq:hubspecificity}
\end{align}
where $p_h$ is the probability that a uniformly sampled FNP is hub-connecting.
As $\lambda_h$ increases from $0$ to $1$, $\mathbb{P}\left(\text{FNP is hub-connecting}\right)$ increases from $p_h$ to $1$.
Theoretically,
\begin{align*}
    p_h = 1-n\left(\frac{|\Omega_h|}{|\Omega|}\right)^2-\left(1 - n \frac{|\Omega_h|}{|\Omega|}\right)^2\,,
\end{align*}
where $|\Omega_h|$ is the area of each hub, $n$ is the number of hubs, and $|\Omega|$ is the area of $\Omega$.
For our investigations ($n = 4, |\Omega_h| = |\Omega|/34$), $p_h \approx 0.22$.
$u$ is then resampled for the next FNP.

\textbf{Acceptance-rejection Test for Rich-Club Specificity constraint. }
To sample an FNP with parameter $\lambda_r$, a value $u$ is first sampled from the $\mathrm{Uniform}[0,1]$ distribution.
If $u < \lambda_r$, the source and target points of the FNP are repetitively sampled uniformly from $\Omega$ and accepted when the FNP is a rich-club connection, i.e., it connects a point inside any one hub with a point inside another hub.
Otherwise, if $u \geq \lambda_r$, the source and target points are sampled once and accepted. 
In this way, the probability that any given FNP is hub-connecting is 
\begin{align}
    \mathbb{P}\left(\text{FNP is a rich-club connection}\right) = (1 - \lambda_r)p_r + \lambda_r, \label{SI:eq:richclubspecificity}
\end{align}
where $p_r$ is the probability that a uniformly sampled FNP is a rich-club connection.
As $\lambda_r$ increases from $0$ to $1$, $\mathbb{P}\left(\text{FNP is a rich-club connection}\right)$ increases from $p_r$ to $1$.
Theoretically,
\begin{align*}
    p_r = n(n-1)\left(\frac{|\Omega_h|}{|\Omega|}\right)^2\,,
\end{align*}
where $|\Omega_h|$ is the area of each hub, $n$ is the number of hubs, and $|\Omega|$ is the area of $\Omega$.
For our investigations ($n = 4, |\Omega_h| = |\Omega|/34$), $p_r\approx 0.01$.
$u$ is then resampled for the next FNP.

\newpage

\bibliographystyle{unsrtnat_et_al}
\bibliography{LocalLRModel}

\end{document}